\documentclass[10pt,journal,comsoc]{IEEEtran}

\usepackage{cite}

\usepackage[T1]{fontenc} 
\usepackage{amsmath}
\usepackage{bm} 

\usepackage{graphicx}
\usepackage{comment}
\usepackage{longtable}
\usepackage{multirow}
\usepackage{rotate}
\usepackage{paralist}
\usepackage{subfigure}
\usepackage{tikz}
\usepackage{array}
\usepackage{xspace}
\usepackage{balance}

\newcommand{\Adv}{Adv}
\newcommand*\circled[1]{\tikz[baseline=(char.base)]{
            \node[shape=circle,draw,inner sep=0.7pt] (char) {#1};}}
\newcommand{\ndnname}[1]{{\footnotesize #1}}
\newcommand{\newpar}[1]{\medskip\noindent{\bf #1.}}

\newcommand{\neb}{Nebula\xspace}

\usepackage{amssymb}
\usepackage{pifont}
\newcommand{\cmark}{\ding{51}}%
\newcommand{\xmark}{\ding{55}}%

\begin{document}

\markboth{M. Ambrosin et al.}{Security and Privacy Analysis of NSF Future
Internet Architectures}

\title{Security and Privacy Analysis of NSF Future Internet Architectures}

\author {
  \IEEEauthorblockN{Moreno Ambrosin\IEEEauthorrefmark{1}, Alberto
    Compagno\IEEEauthorrefmark{2}, Mauro Conti\IEEEauthorrefmark{1}, Cesar Ghali\IEEEauthorrefmark{3}, Gene Tsudik\IEEEauthorrefmark{3}\\} 
  \IEEEauthorblockA{\IEEEauthorrefmark{1}University of Padua -- email: \{ambrosin, conti\}@math.unipd.it\\}  
  \IEEEauthorblockA{\IEEEauthorrefmark{2}Sapienza, University of Rome -- email: compagno@di.uniroma1.it\\}  
  \IEEEauthorblockA{\IEEEauthorrefmark{3}University of California Irvine --
    email: \{cghali, gene.tsudik\}@udi.edu}
}

\maketitle

\begin{abstract}
The Internet Protocol (IP) is the lifeblood of the modern Internet. Its simplicity and universality 
have fueled the unprecedented and lasting global success of the current Internet. 
Nonetheless, some limitations of IP have been emerging in recent years. 
Its original design envisaged supporting perhaps tens of thousands of static 
hosts operating in a friendly academic-like setting, mainly in order to facilitate email 
communication and remote access to scarce computing resources. At present IP interconnects 
billions of static and mobile devices (ranging from supercomputers to IoT gadgets)  
with a large and dynamic set of popular applications. Starting in mid-1990s,
the advent of mobility, wirelessness and the web substantially shifted Internet usage and
communication paradigms. This accentuated long-term concerns about the current
Internet architecture and prompted interest in alternative designs.

The U.S. National Science Foundation (NSF) has been one of the key supporters of 
efforts to design a set of candidate next-generation Internet architectures. 
As a prominent design requirement, NSF emphasized ``security and privacy by design''
in order to avoid the long and unhappy history of incremental patching and retrofitting 
that characterizes the current Internet architecture. To this end, as a result of a  
competitive process, four prominent research projects were funded by the NSF in 2010: 
Nebula, Named-Data Networking (NDN), MobilityFirst (MF), and Expressive Internet
Architecture (XIA). This paper provides a comprehensive and neutral analysis of salient
security and privacy features (and issues) in these NSF-funded Future Internet
Architectures. It also compares the four candidate designs with the current IP-based
architecture and discusses similarities, differences, and possible improvements.
\end{abstract}


\section{Introduction}
\IEEEPARstart{T}{he} original Internet was intended to support thousands of
users, mainly in North America, accessing shared resources via dumb
terminals. Nowadays, the Internet connects over 3 billion of mobile and desktop
devices with a variety of applications ranging from simple web browsing to video
conferencing and content distribution. These extreme changes in Internet usage
accentuated limitations of the current IP-based architecture and prompted
research into alternative internetworking architectures.

In 2010, the National Science Foundation (NSF) launched its ``Future Internet
Architecture'' (FIA) program \cite{NSF-FIA}. Originally, FIA was a 5-year
program with the goal of designing a set of candidate next-generation Internet
architectures. In 2015, NSF renewed its commitment with a follow-on ``Future
Internet Architecture -- Next Phase'' (FIA-NP) program. Unlike FIA which focused
on architectural research, FIA-NP emphasizes evaluation, via prototypes,
testbeds, trial deployments, and extensive experimentation.

\begin{figure}[h!t]
	\centering
	\includegraphics[width=\columnwidth]{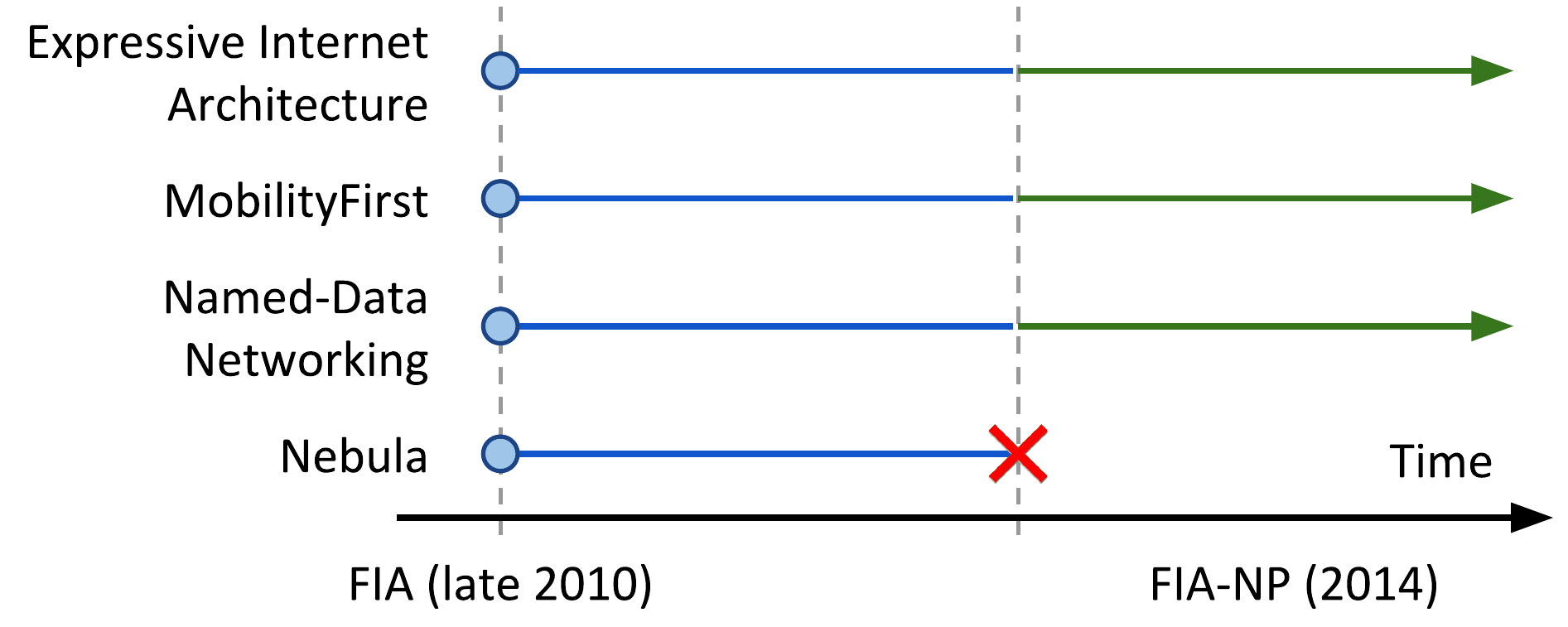}
	\caption{Timeline of FIA \& FIA-NP programs}
	\label{fig:fia_programs}
\end{figure}

FIA originally included four research projects:
Nebula~\cite{nebulaLCNS,Anderson:2014:BON:2656877.2656889}, Named-Data
Networking (NDN)~\cite{zhang2010named}, MobilityFirst (MF)~\cite{mobilityfirst},
and eXpressive Internet Architecture (XIA)~\cite{xia}.
Each project focuses on a new Internet architecture with a distinct vision and
design principles. Nebula envisions a highly-available and extensible core
network interconnecting numerous data centers that enable new means of
distributed communication and computing. NDN focuses on scalable and efficient
data distribution -- thus addressing inadequacies of the current Internet's
host-centric design -- by naming data instead of its location. MobilityFirst
concentrates on scalable and ubiquitous mobility and wireless
connections. Meanwhile, XIA stresses flexibility and addresses the need to
support different communication models by creating a single network that offers
inherent support for communication between various principals (including hosts,
content and services) while remaining extensible to future ones. Only three of
the original four FIA architectures were selected for continued funding under
FIA-NP: NDN, MobilityFirst and XIA. Figure \ref{fig:fia_programs} illustrates
the timeline of each project in FIA and FIA-NP programs.

From the very beginning, one of FIA's key goals (and requirements) is strong
support for security and privacy in all future Internet architectures.
Given the rocky history of security and privacy in the current Internet, this
goal is both very sensible and extremely important.

In this paper, we survey and evaluate security and privacy features in
aforementioned four FIA projects. In doing so, we consider the network layer of
the current Internet architecture as a point of reference. We also show how each
FIA architecture succeeds, exceeds or fails with respect to security and privacy
features of current Internet's network layer, i.e., Internet Protocol (IP)
\cite{postel1981rfc791} and IP Security Extensions (IPSec)
\cite{seo2005security}. We also discuss potential vulnerabilities that can be
exploited to attack transmission channels, end-nodes, and the network
infrastructure. Since different types of resolution services are needed in all
FIA architectures, we compare such security and privacy features of such
services to those of Domain Name System (DNS) \cite{mockapetris2004rfc} and DNS
Security Extensions (DNSSEC) \cite{arends2005dns}.

To the best of our knowledge, this paper represents the first comprehensive
security and privacy treatment of four FIA architectures. Since it is impossible
to predict which FIA architecture(s), if any, will ultimately succeed, we strive
to remain neutral, i.e., to provide a complete and fair analysis of security and
privacy of these architectures.

\newpar{Prior FIA surveys} An early article by Pan et al.~\cite{pan2011survey}
overviews Global Environment for Network Innovations (GENI). Unlike this paper,
\cite{pan2011survey} provides a general overview and does not dwell on security
and privacy aspects.  The work in~\cite{6975502} provides a security analysis of
the four NSF-founded FIA architectures, plus Recursive InterNetwork Architecture
(RINA), Service Oriented Network Architecture (SONATE), and Netlet-based Node
Architecture (NENA).  The analysis in~\cite{6975502} considers four security
features: confidentiality, integrity, availability and authentication. Our work
focuses on the four NSF-funded FIA architectures, with the aim of providing a
more detailed investigation and comparison of a larger set of security
features. Futhermore, different from~\cite{6975502}, we analyze the security of
NSF-funded FIA architecture resolution services.

Other (more focused) surveys addressed security and privacy aspects of
Information-Centric Networking (ICN) architectures
\cite{DBLP:journals/corr/Lutz16,aamir2015denial,abdallah2015survey,
tourani2016security}. Also, \cite{DBLP:journals/corr/Lutz16} analyzed security
and privacy of NDN without comparison with the current Internet
architecture. \cite{aamir2015denial} investigates denial of service attacks in
NDN, and \cite{abdallah2015survey,tourani2016security} provide a more complete
analysis of security in various ICN architectures. It focuses on various
attacks: Naming Related Attacks, Routing Related Attacks, Caching Related
Attacks and Miscellaneous attacks. Instead, \cite{tourani2016security}
distinguishes between security, privacy and access control in ICN.

Other surveys in the literature that focus on ICN architectures do not primarily
focus on security and privacy aspects
\cite{ahlgren2012survey,bari2012survey,tyson2012survey,tyson2013survey}.
\cite{ahlgren2012survey} analyzes Data-Oriented Network Architecture
(DONA), Named Data networking, Publish-Subscribe Internet Routing Paradigm
(PSIRP) and Network of Information (NetInf). It
focuses on naming, routing and forwarding, and caching and mobility. Moreover,
\cite{ahlgren2012survey} marginally considers security and privacy aspects.
Bari et al. \cite{bari2012survey} compare the different naming and routing scheme
adopted by DONA, NetInf, PURSUIT and PSIRP architectures.
\cite{tyson2012survey,tyson2013survey} compare the mobility features of NDN,
DONA, NetInf, PURSUIT. None of
\cite{bari2012survey,tyson2012survey,tyson2013survey} discusses security and
privacy features.

\newpar{Organization} We begin by overviewing IP, IPsec, DNS and DNSSEC in
Section \ref{sec:today-internet}. Next, sections \ref{sec:ndn} through
\ref{sec:xia} summarize \neb, NDN, MobilityFirst and XIA, respectively. Section
\ref{sec:network_sec_and_priv} evaluates security and privacy features of these
architectures, and compares them with those of IP and IPsec. Section
\ref{sec:resolution_service} analyzes security and privacy of resolution
services used by each new architecture. Section~\ref{sec:summary} summarizes our
comparative analysis, and higlights open issues, and possible future research
directions. Finally, Section~\ref{sec:conclusions} concludes our paper.

\section{The Internet of Today}\label{sec:today-internet}
Today's Internet architecture 
was designed over three decades ago to seamlessly inter-connect multiple
heterogeneous networks. At the core of today's Interenet is the TCP/IP protocol
suite, which puts together protocols, applications and network mediums, and
organizes them into four abstraction layers: Link, Internet, Transport and
Application. This design leads to an hourglass shape with IP as the network
layer as its ``thin waist'' \cite{akhshabi2011evolution}.

We consider IP, which operates at the Internet layer, to be our point of
reference when analyzing security and privacy of FIA architectures. IP is
responsible for forwarding packets (a.k.a. datagrams) from the source IP
interface to its destination counterpart. A host may have one or more IP
interfaces, while a router has at least two. Each IP interface is identified by
a distinct fixed-length IP address.

Another fundamental component of today's Internet architecture, and subject of
our analysis, is DNS. DNS is a distributed service that translates
application-specific domain names (specified in URLs) into their corresponding
IP addresses, allowing hosts to communicate using meaningful names, rather than
IP addresses.

In what follows, we briefly describe IP and DNS.

\subsection{Internet Protocol}\label{sec:ip}
The cornerstone of IP is addressing of network devices. Every network-layer
entity (router or host) is identified by at least one IP address which consists
of a network prefix and a host identifier. The boundary between them is
flexible, which allows IP addressing to scale.

An IP datagram contains source and destination addresses along with other fields
that convey control information. Actual data is carried in the payload field.
When a packet is received, a router searches its Forwarding Information Base
(FIB) to identify the next hop for that packet.  A FIB contains a set of
entries, each mapping one or more network prefixes to a router's interface and a
next-hop IP address. This allows routers to perform longest-prefix matching on
the destination address to identify the next hop. If a packet can not be
forwarded, it is dropped and an error message is generated via the Internet
Control Message Protocol (ICMP) \cite{postel1981rfc792}.\footnote{ICMP is also
used for sending control messages, such as routing redirect for networks and
hosts.}

One important IPv4 feature is packet fragmentation. If the size of an IP packet
is larger than the forwarding interface's Maximum Transmission Unit (MTU)
\cite{braden1989requirements}, the packet must be divided into smaller chunks,
called fragments. A destination host must reassemble fragments to recover the
original packet. Other network entities, such as Network Address Translation
tables (NATs) \cite{hain2000architectural,nat_rfc3022}
and firewalls {\em might} also assemble fragments.

As mentioned earlier, IP was originally designed for a small and realtively
amicable research community. Neither its longevity nor its popularity was
foreseen. Thus, it is unsurprsing that IP lacks any security and privacy
features. In late 1980-s and early 1990-s, as IP started to gain global
popularity and the Internet transcended into the commercial sector, IPsec
suite~\cite{seo2005security} was designed to provide basic security services,
such as: origin authentication, data integrity, and confidentiality for IP
datagrams. The first two are attained via Authentication Header (AH) protocol
\cite{kent2005ip_ah}, while Encapsulation Security Payload (ESP) protocol
\cite{kent2005ip_esp} provides all three security features. IPsec supports two
modes of operation:
\begin{itemize}
\item {\em Transport}: provides end-to-end communication, e.g.,
  client-server communications. Only packet payloads are encrypted and
  authenticated in transport mode. Transport and application layers of packets
  are secured by a hash, thus, they can not be modified, e.g., using NAT.
  NAT-Traversal (NAT-T) \cite{kivinen2005negotiation} is developed to overcome
  this issue.
\item {\em Tunnel}: typically used between gateways to provide a secure
  connection (pipe) between physically separate networks, e.g., different sites
  of the same organization. Tunnel mode also supports secure host-to-gateway
  communication. An IP packet is encrypted in its entirety and encapsulated as a
  datagram with a new outer IP header. One popular application of tunnel mode is
  Virtual Private Networks (VPN) \cite{mason2004ccsp}.
\end{itemize}
IPv6 \cite{deering1998internet} is a newer version of IP developed to overcome
some limitations of IPv4. One of its main new features is extended 128-bit
address space (as opposed to 32 bits in IPv4). Another departure from IPv4 is
lack of in-network fragmentation. Before sending an IP datagram must first
discover the smallest MTU on the path to the destination and fragment the
datagram accordingly. To help with this, the Path MTU Discovery protocol
\cite{mccann1996path} was designed and implemented. IPv6 also takes into
consideration security and privacy by implementing some features similar to
IPsec -- such as AH and ESP -- as extension headers \cite{ipv6extensionheaders}.

In the rest of this paper, we use the term ``IP'' to refer to both IPv4 and IPv6,
unless otherwise specified.

\subsection{Domain Name System}\label{sec:dns}
The purpose of DNS is translation of domain names (e.g., those found in URL
prefixes) into IP addresses. Domain names are organized in a hierarchical
fashion: a top-level domain (e.g., ``{\tt .com}'') is followed by many sub-level
domains (e.g. second-level domain ``{\tt example.com}'', and third-level domain
``{\tt sub.example.com}''). For each domain, DNS assigns an {\em authoritative
name} server that stores information of, and responds to queries for, a specific
contiguous portion of the domain name space, called {\em DNS zone}. This
information is contained in {\em Resource Records} (RR-s) -- basic DNS elements
which are also carried in DNS replies. Moreover, authoritative name servers
might delegate authority over sub-domains to other name servers, thus increasing
DNS's scalability.

A user interacts with DNS by issuing a query to a local {\em resolver}: a
process running on the end-user's device which forwards the query to the
appropriate name server(s). The resolver sends a UDP (User Datagram Protocol
\cite{postel1980user}) packet containing the query to the DNS server, which is
usually located in the resolver's local network. The server then checks if it
can reply to the query from its cache. Otherwise, it fetches the response from
other local or remote DNS servers.

DNS queries can be of two types: iterative or recursive. An iterative query
allows a DNS server to return the best answer to the resolver, based on its
local information, i.e., either a cached RR or an RR belonging to its zone. If
the server does not have an exact match for the queried name, it returns a {\it
referral}: a pointer to a DNS server authoritative for a lower level of the
domain namespace. The resolver then queries the DNS server in the referral which
can also reply with a referral. This process continues until the resolver
receives requested information, or an error is generated. In the recursive
query, DNS servers reply with either the requested RR or an error. If the DNS
server does not have the requested information, it recursively queries other DNS
servers.

Although DNS was originally designed as a static distributed database, it now
allows dynamic records updates \cite{bound1997dynamic,wellington2000secure} and
zone transfers~\cite{lewis2010dns}. Also, a recent proposal envisions DNS as a
distributed database to store IP related information
\cite{abley2010nameservers}. For instance, \cite{richardson2005method} proposes
storing IPsec keys related information in DNS records and mapping them to IP
addresses.

The original DNS did not include any security or privacy features. DNS Security
Extensions (DNSSEC) was added to provide data integrity and origin
authentication for DNS messages. In DNSSEC, RR-s are signed by their
authoritative servers' keys. The basic mechanism and the query-response protocol
of DNS remain unaltered.

\section{\neb}\label{sec:nebula}
\neb \cite{nebulaFiaProject,nebulaLCNS,Anderson:2014:BON:2656877.2656889}
is a FIA project focused on providing a secure and cloud-oriented networking
infrastructure. Its architecture is composed of three tiers:
\begin{itemize}
\item {\em Network core} (NCore) is a collection of routers and interconnections
  that provide reliable connectivity between routers and data centers. NCore is
  based on high-performance core routers and rich interconnected topologies
  \cite{180325}.
\item {\em \neb Virtual and Extensible Networking Techniques} (NVENT) represents
  the control plane of \neb. NVENT helps in establishing trustworthy routes based
  on policy routing \cite{Arye:2012:INR:2412096.2412099} and service naming
  \cite{180580}.
\item {\em \neb Data Plane} (NDP) is responsible for routing packets along the
  paths established by NVENT. To guarantee confidentiality, availability, and
  integrity, NDP ensures that packets for a specific communication can only be
  carried when all parties, i.e., end nodes and routers in between, have agreed
  to participate.
\end{itemize}

\subsection{\neb Network Layer}\label{sec:nebula_data_plane}
The original design of \neb specifies different candidate network layer stacks
for NDP \cite{nebulaLCNS}, e.g., ICING \cite{Naous:2011:VEN:2079296.2079326},
TorIP \cite{Liu:2011:TII:2070562.2070576}, and Transit-as-a-Service (TaaS)
\cite{Peter:2014:OTE:2619239.2626318}. From this list, ICING was picked as the
most suitable candidate and was included in the Zodiac \neb prototype
implementation \cite{Anderson:2014:BON:2656877.2656889}.

\begin{figure*}[t]
\centering
\subfigure[Packet forwarding in ICING.]
{
  \centering
  \includegraphics[width=0.775\columnwidth]{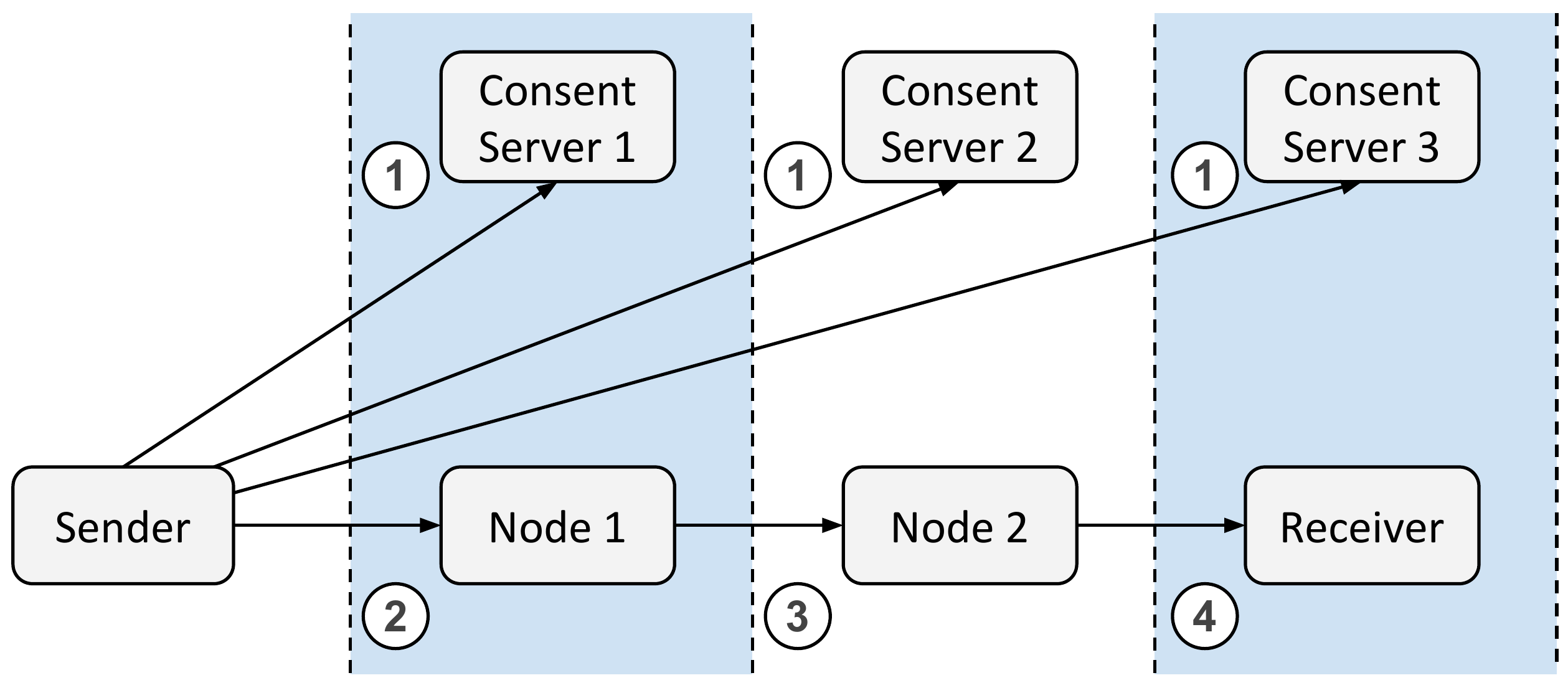}
  \label{fig:forwarding_ICING}
}
\subfigure[ICING packet high-level structure.]
{
  \centering
  \includegraphics[width=1.15\columnwidth]{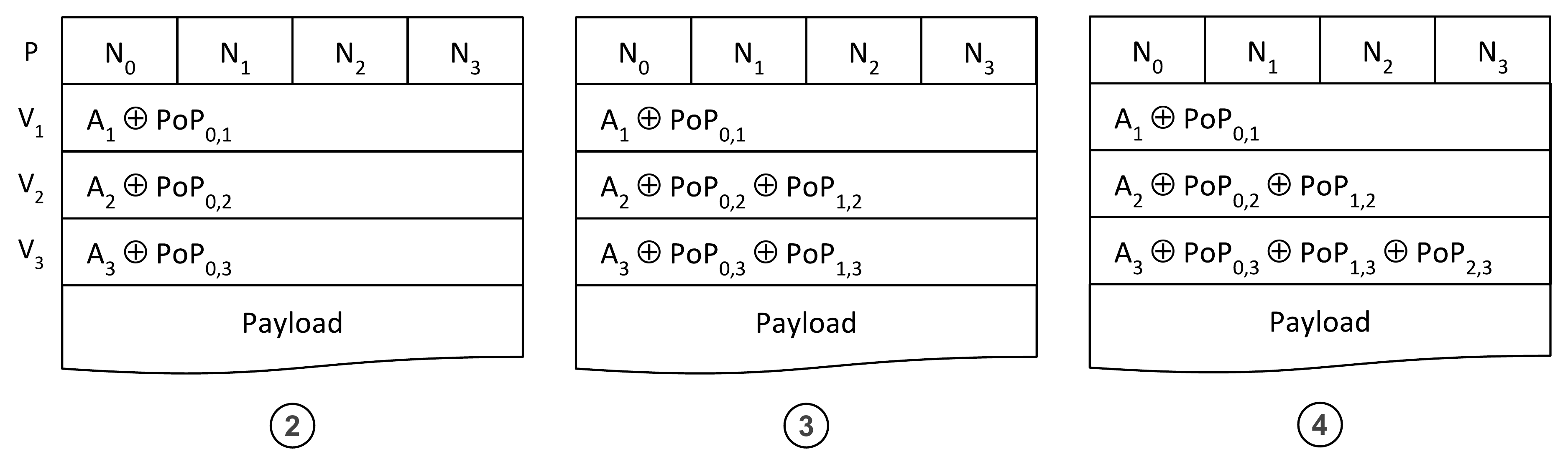}
  \label{fig:packet_ICING}
}
\caption{ICING architecture.}
\label{fig:icing_architecture}
\end{figure*}

ICING provides a new primitive, called {\em Path Verification Mechanism} (PVM),
which guarantees the following two properties:
\begin{itemize}
\item {\em Path Consent} -- every entity in a path between two hosts
{\em consents} the use of the whole path before the communication starts.
\item {\em Path Compliance} -- the possibility for each node in a path between
two hosts to verify that a received packet: (1) follows the approved path; and
(2) has been ``correctly'' forwarded by all the previous nodes in the path, i.e.,
according to a specific pre-established policy.
\end{itemize}
ICING can be deployed either at the network layer or as an overlay on top of IP.
In the former case, service providers can deploy ICING nodes as ingress gateways
to their networks. However, in the latter case, ICING nodes may become
{\em waypoints}, interconnected using IP, providing waypoint-level path
guarantees.

To start communication, a sender must first establish a complete path. Such a
path can be provided by DNS with policy enforcement
\cite{Naous:2011:VEN:2079296.2079326}. Figures \ref{fig:forwarding_ICING} and
\ref{fig:packet_ICING} show how forwarding works in ICING and a high-level
representation of how the ICING header evolves.

Once a path is selected, the sender requests a {\em Proof of Consent} (PoC$_j$),
for each node $j$ on the path (action \circled{1} in Figure
\ref{fig:forwarding_ICING}). PoCs are cryptographic tokens created
by each node transit provider, which attest to the provider's consent to carry
packets along the specified path. Each PoC certifies that the corresponding
network provider consents to (1) the full path, and (2) a specific policy-based
set of local actions (e.g., forwarding) to be performed on packets traversing the
path. PoCs are generated by a {\em consent server}, which is owned by the transit
provider or acts on its behalf. Such servers share secret keys with each node
(router) in their corresponding providers. Once all PoCs are received, the path
is established and packet transmission can begin.

Each packet contains a header (shown in Figure \ref{fig:packet_ICING}) including:
(1) the path $P$ consisting of all ICING nodes  $N_j$ forming it, and, (2) a list
of verifiers $V_j$, one per node $N_j$ in the path except the sender. This
allows each verifier to prove that the packet passed through all previous nodes.

A sender builds a packet header as follows:
\begin{enumerate}
\item {\em Proof of Provenance} (PoP) token, one for each node on the path
  (action \circled{2} in Figure \ref{fig:packet_ICING}), is
  generated using a PoP key $k_j$ shared with the corresponding node $j$. In
  Figure \ref{fig:packet_ICING}, PoPs are denoted as PoP$_{i, j}$,
  where $i$ is the index of the node generating the PoP and $j$ is the index of
  the node for which PoP is generated. Specifically, PoP$_{0, j}$ is computed by
  node $0$ using $k_j$, path $P$, and message $M$ itself.
\item Authenticator $A_j$ is computed for each node $j$ using PoC$_j$, $P$ and
  $M$.
\item Verifiers $V_j$, one per node, are computed by XORing the corresponding
  $A_j$ and PoP$_{i,j}$.
\end{enumerate}
PoP tokens are used by each node on the path to prove that downstream nodes have
handled the received packets based on the established policies. When an
intermediate node $N_i$ receives a packet, it performs the following actions:
\begin{enumerate}
\item Computes the corresponding PoC$_i$.
\item Computes PoP$_{j, i}$ using $k_j$, for each downstream node $N_j$.
\item Verifies that the received PoC$_i$ and PoP$_{j, i}$ match the two
  values computed in the previous two steps. If this verification fails, $N_i$
  drops the packet.
\item Derives a shared PoP key $k_l$, for each upstream node $N_l$, and computes
  PoP$_{i,l}$ as described above.
\item Modifies the verifiers to include the computed PoP, and forwards the packet
  upstream (actions \circled{3} and \circled{4} in Figures
  \ref{fig:forwarding_ICING} and \ref{fig:packet_ICING}).
\end{enumerate}
The previous steps allow any node to guarantee that all packets are forwarded by
all the consenting nodes while establishing the path.

\subsection{\neb Control Plane}
The control plane in \neb is provided by NVENT. NVENT uses declarative networking
\cite{loo2006declarative,loo2009declarative}, and allows administrators to
provide high-level specifications of their routing policies. NVENT also involves
special interfaces, called {\em service interfaces}, that enable service access
and specify the required level of availability. For instance, an emergency
service can request high availability, which can be provided by multi-path
interdomain routing. A distributed resolution service is used for discovery of
other NVENT services. This service is populated by service providers, e.g., NCore
data centers \cite{anderson2010nebula}.

Serval is an implementation of NVENT based on the concept of service-centric
networking \cite{mckinney1998service,griffin2015service}, which decouples
service instances (e.g., web or email services) from their physical locations
(i.e., IP address and port). Serval introduces a new layer, the Service Access
Layer (SAL), between the network layer and the transport layer. With Serval,
each service is identified by a {\em serviceID}, a unique identifier that
applications use to communicate with the service. In addition, each local
traffic flow, representing a connection between two hosts, is identified by a
unique {\em flowID}. The request is handled by SAL, which uses local control
plane policies to map the {\em serviceID} to a service instance. SAL
eventually creates a new {\em flowID} that identifies the established
connection. This {\em flowID} is delivered to the destination host during
connection setup, and used by both parties for connection identification.
Finally, SAL routes the packet based on specific control plane rules contained
in its {\em SAL table}. For instance, a host application that wants to connect
to a specific service might direct the first request to a default Serval
router (using its IP address). The SAL of the router then processes the
request and take further decisions based on its {\em SAL table} (e.g., forward
to another router or send directly to a known service instance). Moreover,
Serval does not directly provide clients a way to learn {\em serviceID}s: It
simply suggests the use of directory services or search engines \cite{180580}.

\begin{figure}[h!]
  \centering
  \includegraphics[width=\columnwidth]{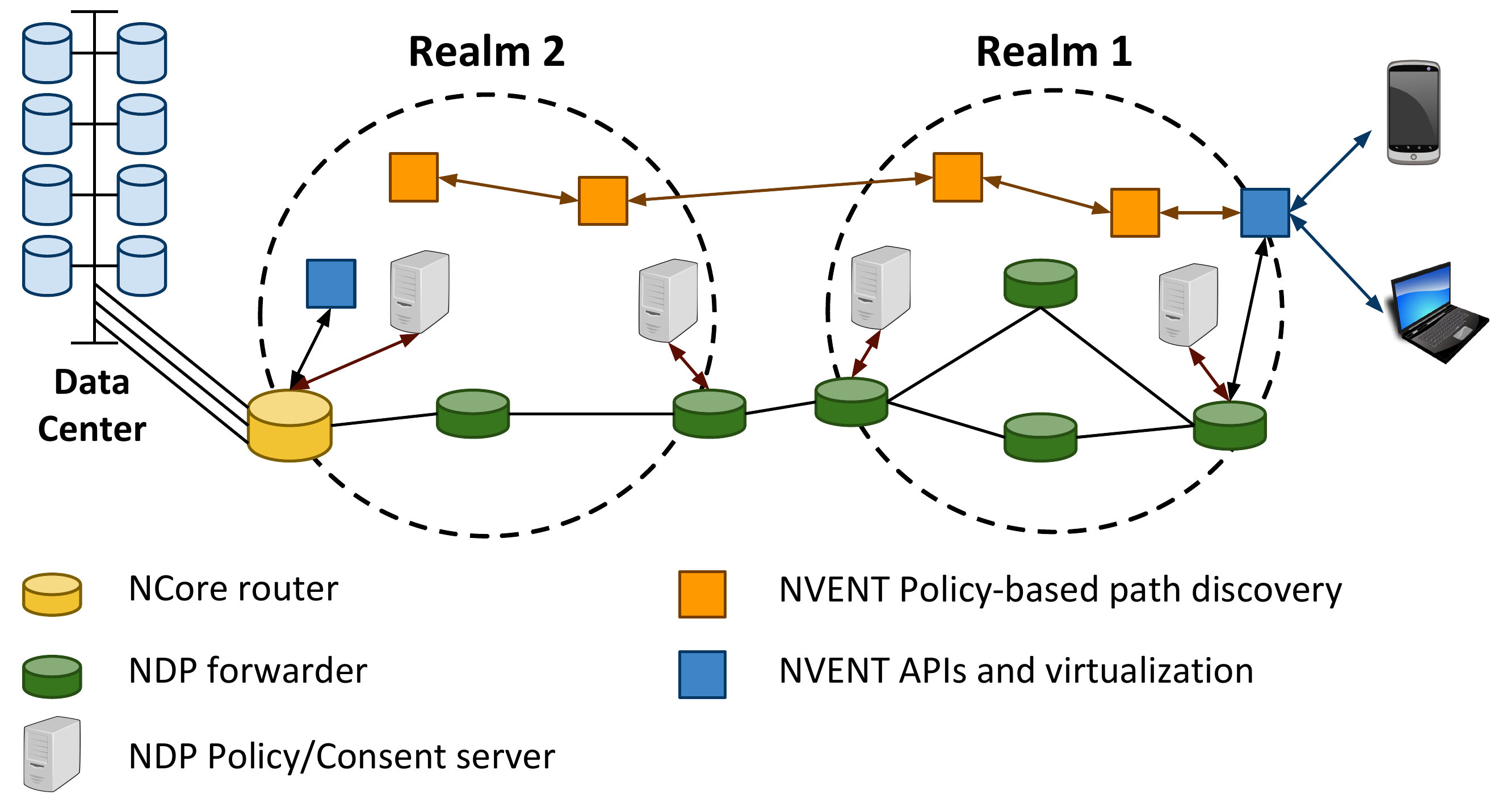}
  \caption{High-level view of \neb components integration}
  \label{fig:nebula_integration}
\end{figure}

Figure \ref{fig:nebula_integration} presents a view of how all \neb
components integrate to allow a user to negotiate a custom end-to-end path to a
specific data center and send the desired packets. First, the user (either the
mobile phone or the laptop in the figure) contacts NVENT to request a path to
NCore. NVENT determines a suitable path that complies with each transit network's
policies and contacts the corresponding consent servers to obtain the necessary
PoCs. Once the path and all PoCs are delivered to the user, the latter generates
appropriate packet headers and forwards them, using the NDP forwarders network,
to the nearest NCore router. This router ensures that all header fields are valid
(as described above) and verifies that the negotiated path has actually been
traversed. Once verified, the core router forwards received packets to the
correct data center using its NCore links.

\section{Named-Data Networking}\label{sec:ndn}
While IP traffic consists of packets sent between communicating end-points, NDN
traffic is comprised of explicit requests for, and responses to, named content
objects. NDN is based on the principle of Content-Centric Networking, where
content, rather than hosts, occupies the central role in the
architecture. NDN is primarily oriented towards efficient large-scale content
distribution. Rather than directly addressing specific hosts, NDN users (called
consumers) request pieces of content by name. The network is in charge of finding
the closest copy of the content, and delivering it. This decoupling of content
and location allows NDN to efficiently implement multicast, content replication
and fault tolerance.

\subsection{NDN Network Layer}
The NDN network layer uses hierarchical structured names to directly
address content. Names are composed of a number of human-readable components,
e.g. \ndnname{/ndn/bbc/frontpage/news} where ``\ndnname{/}'' represents the
boundary between name components. Since names are opaque to the network, they can
contain binary components.

To support content distribution, NDN defines two types of packets: interest and
content (the latter is also called data packet). NDN communication adhered to
the {\em pull} model, that is: every content is delivered to consumers only upon
explicit request. Specifically, a consumer issues an
{\em interest} packet carrying the name of the desired content. The network will
then forward the interest towards the producer.

One important feature of NDN is in-network caching: any router can store a copy
of the content it receives or forwards, and use it to satisfy subsequent
interests. Therefore, an NDN interest might be satisfied by the actual content
producer or any intermediate router.
Along with in-network caching, NDN introduces another important
feature called interest collapsing: only the first of multiple closely
spaced (and timed) interests requesting the same content is forwarded by each
router.

Each NDN entity (not only routers) maintains the following three components
\cite{zhang2010named}:
\begin{compactitem}
\item {\em Content Store} (CS) -- cache used for content caching and retrieval.
A router's cache size is determined by local resource availability. Each router
unilaterally determines what content to cache and for how long. From here on, we
use the terms {\em CS} and {\em cache} interchangeably.
\item {\em Forwarding Interest Base} (FIB) -- table of name prefixes and
corresponding outgoing interfaces. FIB is used to route interests based on
longest-prefix matching of their names.
\item {\em Pending Interest Table} (PIT) -- table of outstanding (pending)
interest names and a set of corresponding incoming interfaces, denoted as
{\em arrival-interfaces}
\end{compactitem}
When an NDN entity receives an interest, it searches its PIT to determine whether
another interest for the same content is pending. There are three possible
outcomes:
\begin{compactenum}
\item If a PIT entry for the same name exists, and the arrival interface of the
present interest is already in {\em arrival-interfaces}, the interest is
discarded.
\item If a PIT entry for the same name exists, yet the arrival interface is new,
the router appends the new incoming interface to {\em arrival-interfaces}, and
the interest is not forwarded further.
\item Otherwise, the router looks up its cache for a matching content. If it
succeeds, the cached content is returned and no new PIT entry is needed.
Conversely, if no matching content is found, the router creates a new PIT entry
and forwards the interest using its FIB. 
\end{compactenum}
Upon receipt of the interest, the producer (or an intermediate router) responds
with a matching content, thus {\em satisfying} the interest. The content is then
forwarded towards the consumer, traversing the reverse path of the preceding
interest. Each router on the path flushes the corresponding PIT entry and
forwards the content out on all interfaces specified by that entry. If a content
is received by a router with no prior matching interest, the content
is considered unsolicited and is discarded. Since no additional information is
needed to deliver content, interests do not carry any form of {\em source
  addresses}.

The last component at the end of content name can carry an implicit digest (hash)
component of the content that is recomputed at every hop. This effectively
provides each content with a unique name.
Names carrying such digest forms what is called as
Self-Certifying Names (SCNs). If an interest is issued using SCN, the retrieved
content is guaranteed, due to longest-prefix matching, to be the same content
requested by the consumer. However, in most cases, the hash component is not
present in interest packets, since NDN does not provide any secure mechanism to
learn a content hash {\em a priori}.

Apart from the name, content packets carry a {\tt Signature} generated by the
content producer and covering the entire content. For this reason, each producer
is required to have at least one public key, represented as a {\em bona fide}
named content object. Other notable fields in content packets are: the
{\tt Payload} containing the actual data of the content and the {\tt ContentType}
defining the type of the content, e.g. data or key. Other important fields in
interest packets are: the {\tt KeyLocator} which references to the public key
required to verify the signature, and the {\tt InterestLifetime} which specifies
the lifetime of an interest before it expires (and its corresponding PIT entry is
flushed).

Similar to IP, fragmentation of NDN packets can not be avoided. The fact that
names can grow arbitrary long might cause interests length to span beyond some
link MTU values. In this case, fragmentation must occur. However, since FIB
forwarding is based on the availability of the entire name, reassembling of
fragmented interests at every hop is a must. Furthermore, interest collapsing
can cause content objects to be fragmented (or even re-fragmented) by routers
\cite{ghali2015secure}. The question remains to whether to perform a hop-by-hop
reassembly \cite{afanasyev2015packet}, or cut-through processing of content
fragments \cite{ghali2015secure}. Regardless of its claimed benefits, it is
trivial to see that hop-by-hop reassembly incurs unnecessary overhead and
end-to-end latency.

Not all interests result in content being returned. If an interest encounters
either: (1) a router that can not forward it further or (2) a producer that has
no matching content, no error is generated. PIT entries in intervening routers
simply expire when no matching content is received. In such case, the consumer
can choose to re-issue the same interest after a timeout.

\subsection{NDNS Distributed Database}
Since content can be addressed using human-readable names, NDN in principle does
not require a resolution service that translate user-friendly names into network
addresses. However, as discussed in \cite{afanasyev2013addressing}, a distributed
database similar to DNS, if existed, provides several benefits to the NDN
architecture:

\begin{itemize}
\item Cryptographic credential management: Since each data packet is required to
be signed, a distributed database is optimal to store and serve security
information (e.g., keys and certificates) for namespaces.
\item Namespace regulation in the global routing: Similar to the ROVER project
\cite{gersch2013rover}, a DNS-like service can store information that certifies
the authorization of ASes to announce a particular prefix in the global routing.
\item Scaling NDN routing: The fact that NDN names can be arbitrary long renders
the namespace infinitely large. This exceeds the number of possible routable IP
prefixes. Therefore, a DNS-like service can be used to implement a Map-n-encap
solution to increase scalability in NDN routing \cite{afanasyev2015map}.
\end{itemize}
Two distributed database systems that resemble the DNS design, KRS and NDNS, are
proposed in \cite{Mahadevan:2014:CKR:2660129.2660154,afanasyev2013addressing},
respectively. These two proposals adopt a similar design and provide the same
features. In the rest of this paper we use NDNS to refer to such a distributed
system.

Similar to domain names in DNS, NDNS organizes namespaces in a hierarchical set
of zones and assigns replicated authoritative servers for each of them. NDNS
queries are expressed via interest, in which, names carry all query's necessary
information. NDNS responses are carried in content objects where their payloads
contain the information requested by the corresponding query.

NDNS reflects many of the DNS protocol machineries: a resolver issues an
iterative or recursive query to a local NDNS server. In case of iterative query,
the server can reply with the answer (if known), a referral, or a negative
response. In case of recursive query, if the NDNS server does not know the
answer, it recursively query another NDNS server until it receives an answer
(i.e., the requested data or a negative response). Moreover, secure dynamic updates
are provided as in DNS~\cite{vixie2000secret}.

\section{MobilityFirst}\label{sec:mobility_first}
MobilityFirst architecture aims to overcome the inefficiencies and limitations
of today's Internet due to mobility. It focuses on scenarios where wireless connections are {\em
ubiquitous} and {\em pervasive}. To this end, MobilityFirst has been designed
around the concepts of \textit{mobility} and \textit{trustworthiness}. All
endpoints must be able to seamlessly switch network connection, and the network
must be resilient to compromised endpoints and routers.

MobilityFirst treats {\em principals} -- devices, content, interfaces,
services, human end-users, or a collection of identifiers -- as primary
addressable network entities. To promote mobility, the (constant) identity of a
principal and its (dynamic) network location are strictly separated. This
requires a distributed Global Name Service (GNS) to bind principal identities to
network addresses. Furthermore, identity and network address separation (1)
facilitates service implementation and deployment, and (2) supports designing
routing protocols that overcome link fluctuation and disconnections
\cite{nelson2011gstar}.

We now briefly describe MobilityFirst's network layer and its Global Name
Service.

\subsection{Network Layer}\label{subsec:MF_network_layer}
Two types of identifiers are used to differentiate between principal identities
and their physical locations.
\begin{itemize}
\item \textbf{Global Unique Identifier (GUID):} a flat self-certifying
  identifier that uniquely identifies a principal. GUIDs can be generated using
  multiple methods depending on the provided service type. For instance, they
  can be derived from the public key of a host or a service principal or the hash
  of a content principal. For the sake of usability, a human readable name can be
  assigned to a principal  and later resolved (by GNS) to the corresponding
  GUID.
\item \textbf{Network Address (NA):} a flat address that identifies a
  \textit{network} to which a particular principal (GUID) is connected.
  MobilityFirst networks are equivalent to ASes on today's Internet. NAs can be
  used to identify finer-grained networks such as subnets or organizations. In
  cases where principals are connected to multiple networks (e.g., using 3G and
  WiFi simultaneously), multiple NAs can correspond to the same principal.
\end{itemize}
As a consequence of this addressing scheme, MobilityFirst defines a new packet
type called Packet Data Unit (PDU). PDUs contain source and destination GUIDs,
lists of source and destination NAs, payload, and other control fields.

In order to communicate with a specific GUID, endpoints need to query GNS to
obtain the corresponding NA. The retrieved tuple (GUID, NA) is then carried in
the PDU header as a routable destination identifier. PDUs are first delivered to
their corresponding destination NAs (using inter-domain routing), and then to the
destination GUIDs (using intra-domain routing). In case of delivery failure, the
packet is stored inside the network (in routers) and GNS is periodically queried
for a new or updated GUID-NA mapping.

Multihoming, anycast, and multicast are supported by multicast GUIDs (MIDs). MID
has the same format as a regular GUID, except its resolution results in a
{\em set of NAs} (instead of, at most, one). Technically, GNS associates one MID
with several GUIDs (the ones belonging to the multicast group). Resolving all of
them results in one or more elements of the output NAs set.

MobilityFirst can also support content distribution networks. In this case, GUIDs
are composed of two parts:
\begin{itemize}
\item Content GUID (CID): uniquely identifies the content and is generated by
  computing the hash of the corresponding content.
\item Publisher GUID (PID): points to the network entity providing the content.
  Such an entity can be the actual content provider, or a third-party content
  repository.
\end{itemize}
A router may be equipped with a cache. This opportunistic caching feature
facilitates content distribution at the network layer by reducing end-to-end
latency and bandwidth consumption. Moreover, MobilityFirst exploits in-network
caching to implement a per-segment (i.e., a continuous set of links with
caching routers at each end) reliable chunk (few hundred of megabytes) transfer.
Each chunk is fragmented and transmitted according to the segment MTU. Then, the
caching router at the other end assembles the entire chunk and stores it. In case
of transferring failure, caching routers can re-transmit a chunk via the same, or
even a different, path.

\subsection{Global Name Service}\label{subsec:MF_GNS}
GNS is an essential part of the MobilityFirst architecture. Its main task is to
map endpoint identifiers (GUIDs or human readable names) to a set of attributes
including the endpoint network address. GNS relies on the following two services:
\begin{itemize}
\item {\em Name Certification Service (NCS):} is equivalent to a Certificate
  Authority (CA). Its purpose is to (1) assign GUIDs to human-readable names and
  (2) attest this mapping by generating certificates. MobilityFirst allows
  multiple NCSs without a global root of trust. Moreover, if GUID space is large
  enough, the need for coordination between different NCSs is eliminated.
\item {\em Global Name Resolution Service (GNRS):} a distributed naming service
  similar to Domain Name System (DNS) that stores the mapping between GUIDs and
  NAs \cite{mukherjeewinlab,liu2013secure,venkataramani2013design}.\footnote{GNRS
  is the actual GNS service that is responsible for maintaining GUID-NA
  mappings.} Two GNRS implementations are evaluated: (1) a distributed hash table
  maintained among all ASes of the Internet (DMap \cite{vu2012dmap}), and (2) a
  number of replica-controllers that migrate data (GUID-NA mappings) between a
  variable number of active replicas (Auspice \cite{sharma2014global}).
\end{itemize}
Regardless of its implementation, GNRS clients interact with the service by
issuing the following requests to the GNRS resolver:
\begin{itemize}
\item \texttt{insert}: register a new GUID-NA mapping when a principal joins the
  network.
\item \texttt{update}: keep the GUID-NA mapping up-to-date when the corresponding
  principal migrates to a new network location.
\item \texttt{query}: retrieve the list of NAs associated with a specific GUID.
\end{itemize}
In \cite{liu2013secure}, a secure version of the above three GNRS request types
is proposed. The secure \texttt{insert} and \texttt{update} requests adopt a
two-step approach to check validity of a GUID-NA mapping. Four network entities
are involved in this process: (1) the user issuing the new GUID-NA mapping, (2)
the local router to which the user is connected, (3) the border gateway router
that connects the user's AS to the rest of the Internet, and (4) the DHCP server
which assigns the user's address.

The user generates and signs the request containing the GUID-NA mapping. Local
and border routers are in charge of verifying validity of the announced mapping.
This is achieved by verifying that the announced NA is the network connected to
the user (and the local router), and querying the DHCP server to ensure that
the returned NA corresponds to the announced GUID. If the NA matches the one
contained in the {\tt update} or the {\tt insert} request, the mapping is
accepted and added or updated in the GNRS table.

In the secure \texttt{query} request, the protocol involves three entities: the
user, the border gateway, and GNRS. The user issues an authenticated request
and the border router checks its validity. The router then forwards the request
to the appropriate GNRS replica. On receipt, the GNRS satisfies the request with
a signed GUID-NA mapping response.

There are several differences \cite{sharma2014global} between GNRS and DNS
\cite{mockapetris2004rfc}. First, GNRS does not restrict the structure of the
names, while DNS only supports hierarchical names. Second, scalability of GNRS
does not rely on TTL-based caching, which has been proven to be ineffective in the
presence of high mobility. Third, GNRS does not statically give the authority to
a replicated server for a specific set of names. Active and on-demand replication
reduce reliance on passive caching and ensure that mapping replicas are always
accessible close to clients.

\section{eXpressive Internet Architecture}\label{sec:xia}
eXpressive Internet Architecture (XIA) is another research effort aiming to
design a new architecture. XIA is based on three types of principals.
{\em Host}-centric networking can support end-to-end communication, such as video
conferencing and file sharing. {\em Service}-centric networking allows users to
access various network services such as printing and data storage services.
Meanwhile, {\em content}-centric networking can support Web browsing and content
distribution. However, XIA's design is extensible in that it can adaptively
provide network evolution and support any new principal type that might emerge in
the future.

A core architectural property of XIA is {\em intrinsic security} of all
principals. Any entity should be able to authenticate the principal it is
communicating with, without trusted third parties. This can be achieved by binding
one or more security properties with principal names. For instance, using the
hash of a service (or a host) public key as its name allows entities to verify
that they are communicating with the desired principal. Similarly, binding
content with its name can be achieved using the hash of the content as its name,
allowing users to verify the integrity of a requested content.

XIA defines three main design requirements:
\begin{enumerate}
\item All network entities must be capable of clearly expressing their intent.
  This is achieved by designing the network to be {\em principal}-centric and
  allowing in-network optimization. Routers can perform principal-specific
  operations when receiving, processing, and forwarding packets.
\item The network must be able to adapt to new types of principals. This is
  essential to support network evolution.
\item Principal identifiers must be intrinsically secure. This depends on the
  principal type, e.g., authenticating hosts in {\em host}-centric networking is
  different than verifying content integrity in {\em content}-centric networking.
\end{enumerate}
Principal identifiers are denoted as XID, where X defines the type of principle.
For instance, HID identifies a host, SID a service, NID a network, and CID a
content.

\subsection{eXpressive Internet Protocol}
In order to comply with the aforementioned requirements, eXpressive Internet
Protocol (XIP) is designed. XIP defines packet format, addressing schemes, and
behavior of all nodes while processing incoming and outgoing packets from/to
various principal types. One of the main features of the XIP addressing scheme is
flexibility of defining multiple (fallback) paths to destinations. This prevents
downtime and service interruption, especially while gradually deploying new
principal types. An XIP address is a directed acyclic graph (DAG) with several
properties:
\begin{itemize}
\item Each address is a single connected component.
\item Each DAG starts with an untyped entry node and ends with one or multiple
  ``sink'' nodes. Thus, each node in the address graph has a unique XID except
  for the entry node.
\item Edges define next hops in the path.
\item Multiple outgoing edges of a single node are processed in the order they
  are listed.
\item Out-degree of each node is upper bounded to restrict performance overhead.
\end{itemize}
\newcommand{\figwidthI}{0.25\textwidth}
\newcommand{\figwidthII}{0.41\textwidth}
\begin{figure*}[t]
\centering
\subfigure[Shortcut routing]
{
	\includegraphics[width=\figwidthI]{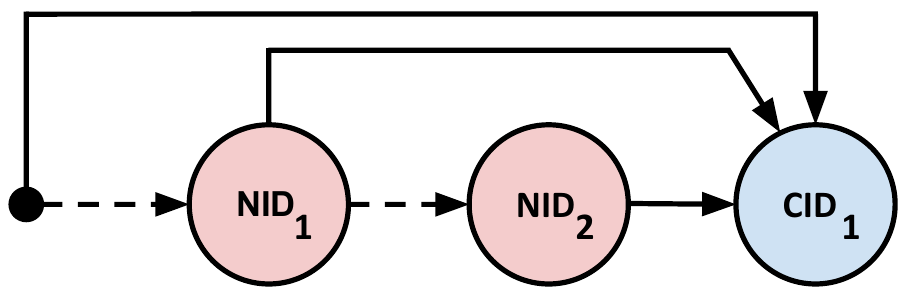}
	\label{fig:xip_shortcut}
}
\subfigure[Binding]
{
	\includegraphics[width=\figwidthI]{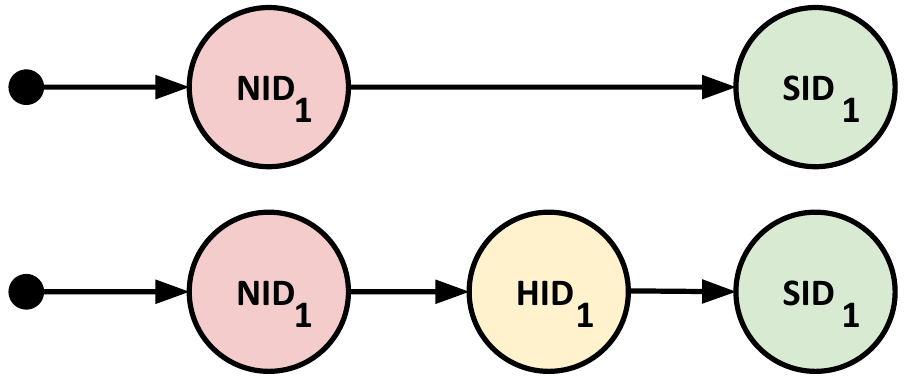}
	\label{fig:xip_binding}
}
\subfigure[Infrastructure evolution]
{
	\includegraphics[width=\figwidthI]{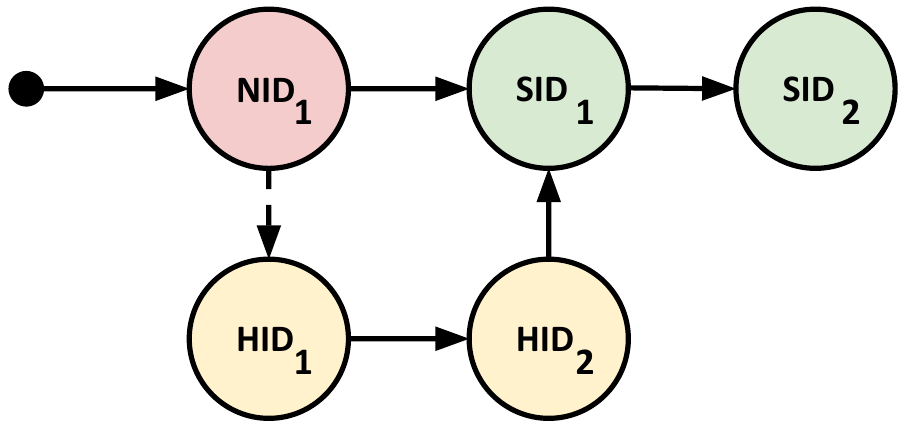}
	\label{fig:xip_evolution}
}\\
\subfigure[Source routing]
{
	\includegraphics[width=\figwidthII]{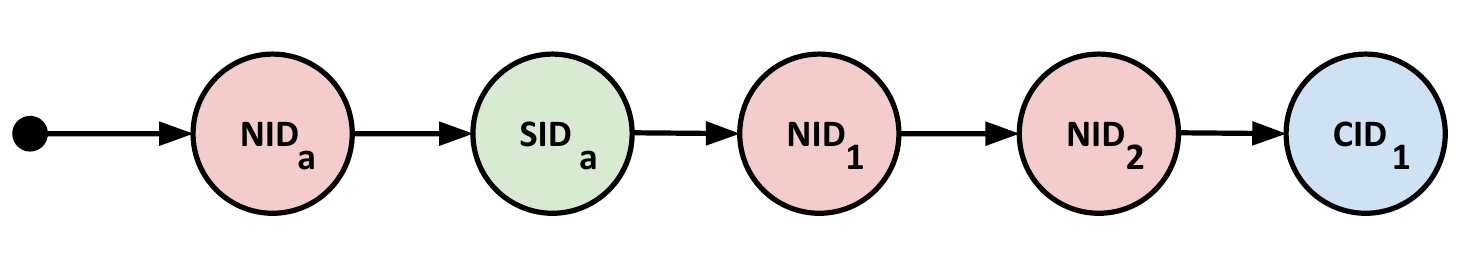}
	\label{fig:xip_source_routing}
}
\subfigure[Multiple paths]
{
	\includegraphics[width=\figwidthI]{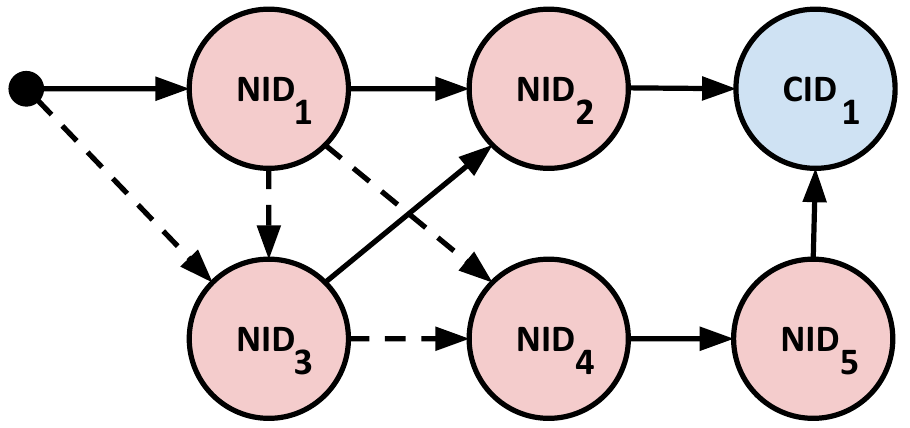}
	\label{fig:xip_multipath}
}
\caption{XIP Addressing Styles}
\label{fig:xip_address_styles}
\end{figure*}
\begin{figure*}[t]
\center
\includegraphics[width=0.75\linewidth]{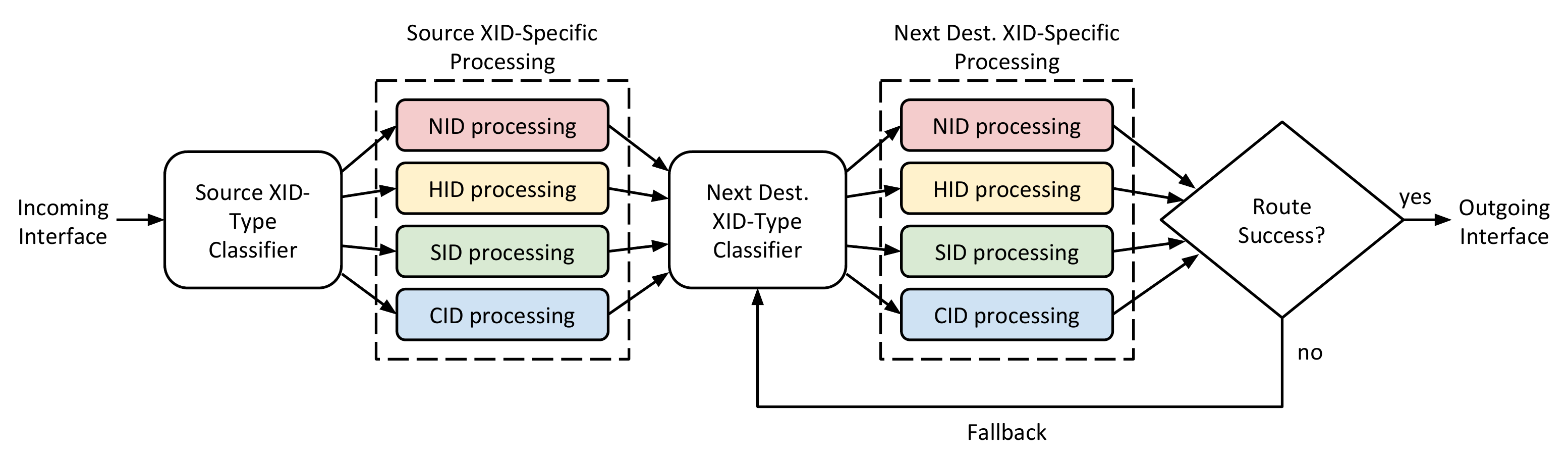}
\caption{XIA Router Diagram}
\label{fig:xia_router}
\end{figure*}
Using DAGs as a basis for XIP addresses allows applications to build several
``styles'' of addresses, such as:
\begin{itemize}
\item \textbf{Shortcut routing} -- this style, shown in Figure
  \ref{fig:xip_shortcut} is best suitable for requesting content
  principal. Each node has a direct edge to the destination principal
  $\mathrm{CID_1}$, which enables in-network caching. If a node does not have the
  content cached, the fallback path is processed and the packet is forwarded to
  the next hop.
\item \textbf{Binding} -- some services require that communication is bound to a
  specific source or destination. For instance, a service hosted in multiple
  geographical locations. Users can establish a session with the closest host
  providing this service. Then, all further communications must be directed to
  this particular host. Figure \ref{fig:xip_binding} shows an
  example of this addressing style. The first packet is destined to
  $\mathrm{SID_1}$, i.e. the closest host, while the second packet is destined to
  $\mathrm{SID_1}$ provided by a specific host $\mathrm{HID_1}$.
\item \textbf{Infrastructure evolution} -- as mentioned above, XIA supports
  gradual network evolution for emerging principal types using fallback
  paths. Figure \ref{fig:xip_evolution} shows an example of this
  style. Assume that $\mathrm{NID_1}$ is gradually deploying service
  $\mathrm{SID_1}$. All $\mathrm{NID_1}$ routers that are not yet updated to
  recognize and process $\mathrm{SID_1}$ use the fallback path through
  $\mathrm{HID_1}$ and $\mathrm{HID_2}$.
\item \textbf{Source routing} -- Figure \ref{fig:xip_source_routing}
  gives an example of this addressing style, in which the source routes the
  packet to the destination through a third party domain and service,
  $\mathrm{NID_a}$ and $\mathrm{SID_a}$, respectively.
\item \textbf{Multiple paths} -- this supports recovery from link failures. An
  example of this style is shown in Figure \ref{fig:xip_multipath}.
\end{itemize}
Figure \ref{fig:xia_router} shows a high level overview of an XIA
router. Its modular design allows efficient multi-principal processing and
supports network evolution. Each router contains two main XID-specific processing
modules:
\begin{itemize}
\item \textbf{Source XID-specific processing:} necessary for certain XID types.
  For instance, in case of a reply to a $\mathrm{CID}$ request, the
  ``$\mathrm{CID}$ processing'' unit can implement in-network content caching.
\item \textbf{Next Destination XID-specific processing:} invoked by the
  {\em Next Destination XID-specific Classifier} which determines the appropriate
  forwarding action. Similar to source processing, this module consists of
  several units that carry on XID-specific operations right before forwarding the
  packet.
\end{itemize}
If all outgoing DAG edges of a node lead to unrecognizable XIDs, the packet is
dropped and an unreachable destination error is generated. It is the
responsibility of user applications to provide appropriate fallback paths to
avoid forwarding failures at any router. Usually fallback paths are built using
well-supported principals, e.g. $\mathrm{HID}$ and $\mathrm{NID}$.

\subsection{Principals}
As mentioned above, principals in XIA support emerging communication paradigm on
the current Internet. When introducing a new principal, the following issues
arise:
\begin{itemize}
\item What does it mean to communicate with a principal of this type?
\item How is the principal's unique XID generated and how does it map to
  intrinsic security properties?
\item What are the source and next destination XID-specific processing actions
  that routers should perform and how can such actions be implemented?
\end{itemize}
We now describe several principal types and discuss their addressing schemes,
in-router processing behaviors, and security properties.

\subsubsection{Network and Host}
Network and host principal identifiers are denoted as $\mathrm{NID}$ and
$\mathrm{HID}$, respectively. They are generated by using the public key
hash of the network or the host. Unlike hosts on current Internet, each XIA
host has a unique $\mathrm{HID}$ regardless of the interface it is communicating
through. This feature helps support host mobility. In order to support fallback
paths, all XIA routers should implement $\mathrm{NID}$ and $\mathrm{HID}$
processing modules.

As mentioned above, the fact that the network and host addresses are derived from
their corresponding public keys allows users to verify the identity of entities
with whom they are communicating. Furthermore, this security requirement helps
defend against address spoofing, Denial of Service (DoS), and cache poisoning
attacks.

\subsubsection{Service}
Services in XIA represent applications in today's Internet. Users communicating
with a service $\mathrm{SID}$ can use a destination address of the form
$\mathrm{NID}$:$\mathrm{HID}$:$\mathrm{SID}$. In today's terminology, this is
analogous to sending a packet to a specific host in a specific network and
indicating the associated protocol and port number.

Since different services might require different specialized processing,
implementing in-router source and next destination processing modules is a
challenge. Therefore, routers are only required to perform default processing,
routing, and forwarding of $\mathrm{SID}$ packets. All other specialized
processing should be handled by end-nodes.

$\mathrm{SID}$s are generated by computing the hash of the service public key.
This inherits security properties similar to $\mathrm{NID}$s and $\mathrm{HID}$s.

\subsubsection{Content principals}
This principal type signifies user's intent to retrieve content. Packets carrying
content identifiers ($\mathrm{CID}$) as destination addresses will be routed all
the way to the node hosting the content. Routers can use a cached version of the
content as a reply to such packets. As mentioned above, caching is implemented by
routers source XID-specific processing module.

$\mathrm{CID}$s are generated based on the cryptographic hash of the content they
address. This binds the content to its name, forming a self-certifying name.

\section{Network-Layer Security and Privacy}\label{sec:network_sec_and_priv}
According to NSF, one of the guiding principles for a new Internet architecture
is to provide security and privacy by design. For this reason, FIA involved
projects do not only focus on improving the network performance, but also on
increasing network security and privacy guarantees. 

In this section, we provide a comparison between the security and privacy
features offered at the network layer of each architecture introduced above, and
compare them with IP and IPsec. We consider the following security and privacy
features that we believe essential to obtain a secure and privacy preserving
communication \cite{shirey2007internet}.

\begin{itemize}
\item {\bf Trust:} A feeling of certainty (sometimes based on inconclusive
  evidence) that an entity will behave exactly as expected.
\item {\bf Data origin authentication:} The corroboration that the source of the
  received data is as claimed.
\item {\bf Peer entity authentication:} The corroboration that a peer entity
  in an association is the one claimed.
\item {\bf Data integrity:} The property that data has not been changed, in an
  unauthorized or accidental manner.
\item {\bf Authorization and access control:} An approval that is granted to
  an entity to access some resources and the protection of such resources against
  unauthorized access.
\item {\bf Accountability:} The property that ensures that entity actions can be
  traced to said entity, which can then be held responsible.
\item {\bf Data confidentiality:} The property that information is not made
  available or disclosed to unauthorized entities.
\item {\bf traffic flow confidentiality:} A confidentiality service to
  protect against traffic analysis.
\item {\bf Anonymous communication:} The guarantees that entity identities being
  unknown or concealed.
\item {\bf Availability:} The property of being accessible and usable upon demand
  by an authorized entity.
\end{itemize}

\subsection{Trust}\label{subsec:trust_model}
IPsec defines trust as a one-way relationship between two or more entities (hosts
or networks). This relationship is represented using a Security Association (SA).
SAs contain a set of information that can be considered as a ``contract'' between
the involved entities. This describes security services and contains
security information needed by hosts to protect the communication.

Entities involved in secure communication in IPsec establish SAs via the ISAKMP
\cite{maughan1998internet} protocol and exchange necessary cryptographic material
using the Internet Key Exchange (IKE) protocol \cite{ikev2_rfc5996}. Host
authentication in ISAKMP and IKE can be achieved via either digital signatures,
or pre-shared keys. Digital signatures require the use of certificates to bind
entity identities to their public keys. This implies the existence of a CA to
create, sign, and properly distribute certificates.

\neb's ICING-based network layer defines trust in a way that is orthogonal to
IPsec. Trust is defined between a host and all nodes forwarding its packets. As
described in Section \ref{sec:nebula}, a host agrees on a ``contract'' with
the network providers carrying the data (i.e., the path negotiated using NVENT)
to specify the operation executed at each hop. Such contracts are
cryptographically enforced. ICING assumes mutual trust between forwarding nodes
and their consent servers, i.e., servers responsible for creating PoC tokens.
Therefore, this notion of trust does not require any PKI
\cite{Naous:2011:VEN:2079296.2079326}. However, ICING does not provide an
end-to-end definition of trust, which can be added by adopting an IPsec-like
approach.

Unlike IP, the notion of trust in NDN is not directly associated with hosts and
networks, but rather with content. Trust in content can be expressed at different
levels of granularity, from a single content to an entire namespace. Recall that
a content object is signed by its producer which allows anyone to verify its
origin and authenticity. Origin verification refers to content producer rather
than whoever stores a copy of that content. In order to authenticate a content
and its origin, its signature must be verified. To do so, the verification
(public) key must be retrieved and trusted.\footnote{Keys in NDN are distributed
in content objects with type {\tt KEY}. Such objects (keys) are signed by their
issuer, e.g., CA.} However, trust management is not specified at the network
layer and is left to applications. NDN network-layer trust management in
discussed in \cite{ghali2014network}.

MobilityFirst places trust in principal. Depending on their type, trust may be
established with (1) hosts, similar to IPsec, (2) content, similar to NDN, or (3)
centralized or distributed services. Trust semantics in XIA also vary depending
on principal types. However, the intrinsic security feature of these principals
(described in Section \ref{sec:xia}) increases trustworthiness of end-to-end
communication and content retrieval. For instance, ensuring that a content hash
matches its identifier allows receivers (and caching routers) to trust that
content.

As shown in Section \ref{sec:xia}, an XIA address consists of a DAG containing a
(partial) path to the destination. To provide trusted path selection for
host-to-host communication, SCION is integrated with XIA \cite{naylor2014xia}.
SCION \cite{zhang2011scion} is an architecture that provides control and
isolation for secure and highly available end-to-end communication. The network
is divided into multiple trust domains consisting of several Autonomous Systems
(ASes) that trust each other. Each domain has a trusted root AS responsible for
relaying packets to and from other domains. Roots initiate path establishment to
all hosts in their domains based on local policies and available bandwidth. This
process results in constructing a path between each host and its domain root.
Whenever two XIA hosts, in different domains, want to communicate, the two half
paths (from each host to its domain root) are combined to establish a complete
end-to-end path. Such path is trusted since it is created by the trusted roots of
each domain.

\subsection{Data origin authentication}\label{sec:data_authenticity}
IP (IPv4 in particular) does not provide any form of authentication.
A separate add-on method, IPsec, provides entity authentication via AH and ESP
protocols.\footnote{Recall that IPv6 implements both AH and ESP as extension
headers.} In transport mode, two hosts securely negotiate a shared secret key.
This key is later used to generate a Message Authentication Code (MAC)
\cite{krawczyk1997hmac} for each packet. Successful MAC verification ensures
authenticity of received packets and their origin. In case of gateway-to-gateway
communication, gateways can only verify that the received data originated by
{\em any} (not a specific) host connected to the network at the other end of the
tunnel. In host-to-gateway communication, the gateway can actually verify that
the data originated by the involved host, while the latter can only verify that
received data is originated by the network located behind the gateway. This
partial authentication opens the door for insider attacks.

\neb's ICING-based network layer does not directly provide data origin
authentication. Instead, it is delegated to applications. ICING allows a
sender to authenticate entities issuing cryptographic tokens, i.e., PoCs.
However, the design does not specify how PoCs are retrieved or authenticated
\cite{Naous:2011:VEN:2079296.2079326}.

NDN provides origin and data authentication via content signatures. Before
consuming content, consumers are required to verify its signature
\cite{zhang2010named}. However, this operation is optional for routers because
signature verification is an expensive operation at line speed and comprehensive
trust management is not viable at the network layer. Even if we assume that
routers know all possible application trust models, establishing trust in content
is complicated and expensive. For instance, traversing a PKI hierarchy requires
routers to fetch and verify public key certificates until a trusted anchor is
reached.

On the other hand, NDN interests can optionally be authenticated using digital
signature \cite{ndn_signed_interests}. In a signed interest, the last component
of the name carries a signature computed by the consumer. Although this reveals
consumer identities, signing signatures (or any other form of identities) are
usually required in some scenarios such as access control.

MobilityFirst and XIA do not provide any data origin and entity authentication at
the network layer. However, their usage of self-certifying identities as
principal identifiers facilitates entity authentication. Recall that for host,
network, and service principals, identifiers are generated by computing the hash
of the public key associated with these principals. Therefore, entity
authentication can be achieved by ensuring that such principal identifiers match
their keys. Peer authentication for content principals can be achieved similar to
NDN since such principals are self-authenticating. Neither MobilityFirst nor XIA
provides a secure mechanism for securely retrieving content identifiers.

\subsection{Peer entity authentication}
\label{subsec:peer-entity-authentication-network}
IPsec provides peer entity authentication during SA establishment of a secure
communication. ISAKMP and IKE, the IPsec's protocols used to establish SAs, can
achieve peer entity authentication using digital signature or pre-shared key.
Digital signatures requires the use of certificates, which bind entity identities
to their public keys. The use of certificates implies the existence of a trusted
third party or a CA to create, sign and properly distribute certificates.
Pre-shared keys on the other hand requires the communicating parties to agree on
the shared secret key before communication begins.

In \neb, ICING allows a sender to authenticate the entities issuing the
necessary cryptographic tokens, i.e., PoCs. However, ICING design does not
specify how PoCs are retrieved, nor does it specify how entities are authenticated
\cite{Naous:2011:VEN:2079296.2079326}.

At its current state, NDN does not provide peer entity authentication
for consumers and producers. However, in case the authentication of one or both
entities is necessary, applications can exploit some features provided by the
network layer. Considering consumers, signed interests can facilitate their
authentication. Whereas for producers, content signature can ensure that the
content is generated by the expected producer. Moreover, if interests
must be satisfied by producers only (and not in-network caches), they
should carry unique names that avoid cache hits and guarantee their delivery to
corresponding producers.

Similar to NDN, MobilityFirst and XIA do not provide peer entity
authentication. However, their usage of self-certifying identities (or names) as
principal identifiers facilitates this task. Recall that for host, network, and
service principals, identifiers are generated by computing the hash of the public
key associated with these principals. Therefore, entity authentication can be
achieved by ensuring that such principal identifiers match their keys. Also,
peer authentication for content principals can be achieved similar to NDN. Note
that neither MobilityFirst nor XIA provides a secure mechanism for securely
retrieving content identifiers.

\subsection{Data integrity}\label{sec:data_integrity}
Although IPv4 header contains the \textit{Header Checksum} field that provides
transmission error detection (a form of integrity check), it does not prevent
packet manipulation. In fact, both versions of IP, introduced in Section
\ref{sec:ip}, completely delegate integrity to IPsec AH and ESP protocols.
Specifically, the HMAC values in these protocol headers are used to achieve
integrity. Depending on the IPsec mode used, host-to-host, gateway-to-gateway, or
host-to-gateway integrity guarantees can be provided by both AH and ESP
protocols. However, AH provides integrity for the entire packet (except for
mutable fields), while ESP guarantees packet headers integrity only.

Each packet in \neb\ carries a sequence of cryptographic verifiers $V_j$, one for
each hop on the path (see Section \ref{sec:nebula_data_plane} for details). The
packet hash is used as part of $V_j$'s calculation. Therefore, ICING guarantees
that neither the packet nor the path can be modified. Also, ICING is recommended
only at domain gateways \cite{Naous:2011:VEN:2079296.2079326}. Thus, integrity can
only be guaranteed by border routers. Within domains, such guarantees are
deferred to either the network-layer protocol or the application.

The way NDN provides integrity is through content signature. By verifying this
signature, consumers and routers can always detect malicious manipulation.
However, when content is requested using SCNs, data integrity is achieved by
comparing the content hash to the last name component of its name. Furthermore,
only signed interests can provide interest integrity.

In both MobilityFirst and XIA, integrity is only available for content principal
types. This is again due to the fact that such a principal identifier is
generated based on the content hash itself. Whenever a content is received, its
hash is compared with its identifier to ensure content integrity. For other
principal types, MobilityFirst and XIA defer integrity guarantees to the
application.

\subsection{Authorization and access control}\label{sec:auth_and_access_control}
Access control in IP is achieved by restricting access based on source and
destination addresses. This is implemented using Access Control Lists (ACLs)
\cite{acl}, which contain a set of rules that grant or deny access to network
resources. When implemented in routers, ACLs specify whether a received packet
will be forwarded further to the next hop, or simply getting dropped. Whereas
host ACLs are used to decide whether to forward packets up the stack towards the
application. Since IP does not natively provide packet integrity, address
spoofing can be used to circumvent ACL rules. Employing IPsec, however,
prevents such actions.

In \neb, paths must be established before communication begins, i.e., clients
must obtain required PoC tokens. Therefore, access control can be implemented by
the consent server granting or denying PoC requests. Traffic sent without valid
PoC tokens can be easily detected and dropped.

Unlike IP, enforcing access control in NDN should be done based on content and
not network entities. Although not implemented in practice, ACLs can still be
used to implement access control. In this case, rules are applied on interest
messages and content objects based on the names they carry. Longest-prefix can
also be employed to grant or restrict access to entire namespaces. Due to the
fact that NDN interests do not carry source addresses, access control on the
consumer granularity can only be achieved in cases where interests are signed
or carry some form of consumer identity \cite{ghali2015interest}.

One way of providing access control in NDN is by using encryption. Producers
can encrypt their content and disseminate decryption keys to authorized consumers
only. Such keys can be encapsulated in content objects and should not be cached.
One drawback of this approach is that it requires consumers to issue at least two
interests for each content (one to request the content itself and one to request
the corresponding key).

Since MobilityFirst and XIA can support different principal types, they
facilitate the combination of both NDN- and IP-based access control schemes. For
content principals, access control is done at the content granularity, similar to
NDN, e.g., content is encrypted using keys disseminated to only authorized users.
For all other principal types, ACLs can restrict access to hosts and other
network services.

\subsection{Accountability}\label{sec:accountability}
One of the main problems in IP is accountability. In fact, IP is subject to
source address spoofing that lead to the inability of tracing back the entity
responsible for a particular action. A simple countermeasure against IP spoofing
requires ASes to implement egress filtering and ensure that all outgoing traffic
carries source addresses owned by these ASes. IPsec guarantees peer entity
authentication when establishing SAs between hosts. Thus, accountability can be
achieved.

\neb provides accountability through path establishment. All routers on a path
consent to use the whole path before the communication begins. Moreover, the fact
that these routers pre-agree on performing a specific set of actions on each
packet passing through allows the detection of any malicious activities.

NDN provides full accountability of producers. Since every
content is signed by its producer, tracing the producer responsible for
generating content is a trivial task. However, accountability can not be provided
if content is served from router caches. Consumers accountability, on the other
hand, can only be achieved when they issue signed interests, or include their
identities in the interests themselves. Otherwise, accountability can not be
provided.

Both MobilityFirst and XIA do not provide accountability at the network layer.
However, signing requests and responses can provide this feature in a similar
fashion to NDN, especially for content principals. Also, IPsec-similar
techniques can be employed to provide accountability for other principal types.

\subsection{Data confidentiality}\label{sec:data_confidentiality}
The natural way to achieve data confidentiality at the network layer is by using
encryption.

IP does not provide data confidentiality. This is done by using IPsec. The level
of confidentiality depends on the mode of operation. In transport mode, ESP only
encrypts the IP packet payload and data confidentiality is host-to-host. Tunnel
mode extends confidentiality to the entire encapsulated IP packet, including both
payload and header. However, data confidentiality can only be achieved in
host-to-gateway or gateway-to-gateway scenarios. Also, ESP confidentiality is not
generally effective against active adversaries. It has been demonstrated that
achieving confidentiality without a strong integrity mechanism, or even applying
integrity before encryption, can only protect against passive adversaries
\cite{bellovin1996problem,krawczyk2001order,4223237}. Thus, even though IPsec
provides confidentiality, poor usage practices can negate its benefits.

\neb's ICING-based network layer does not natively provide data confidentiality.
Instead, it can be achieved by combining ICING with end-to-end encryption of the
packet payload.

Data confidentiality in NDN can be attained by encrypting content payload. This
is not supported by the architecture and is left to the application. However,
even if content is encrypted, the fact that it carries a human-readable name
might leak information about its data.

MobilityFirst and XIA provides content principal confidentiality using methods
similar to the those used in NDN. Fortunately, and due to the fact that such
principal identifiers are generated using the hash of the content itself,
inspecting them does not leak information about the encrypted content. Moreover,
confidentiality of data communicated between host, network and service principals
can be achieved using similar techniques to IPsec.

\subsection{Traffic flow confidentiality}\label{sec:traffic_flow_confidentiality}
It is well known that encryption does not protect against statistical traffic
analysis -- attacks that monitor traffic in order to extract properties, such as
volume and timing \cite{raymond2001traffic}.

IPsec provides some traffic flow confidentiality by padding packet payloads to
hide their size patterns. However, according to IPsec specifications, this is not
mandatory and, therefore, may not be supported in all IPsec implementations
\cite{seo2005security}.

NDN, MobilityFirst, \neb and XIA are all susceptible to traffic analysis attacks
Fortunately, padding can be used to provide traffic and flow confidentiality.

Another architecture-agnostic alternative is to add artificial delays to
communications to better hinder time-based attacks. This, however, comes at the
expense of increasing end-to-end latency and reducing overall network
performance, especially for time-sensitive traffic.

\subsection{Anonymous Communication}\label{sec:anonymous_communication}
IP (with or without IPsec) does not support anonymous communication. This is
mainly because source and destination addresses are in the clear in packet
headers. However, partial anonymity can be achieved using the tunnel mode of
IPsec along with ESP. This is because tunnel mode allows the ESP
protocol to encrypt the original IP packet along with the source and destination
addresses, and it encapsulates that packet into a new one with a
new header reflecting gateway addresses. This combination hides end-host
identities among the set of other hosts connected to respective end-networks.
However, this is only effective if the adversary is eavesdropping on the link
between the two gateways and is not located inside one of the end-networks.
Furthermore, in case of host-to-gateway tunnel mode, only anonymity of hosts
located behind the gateway is preserved.

Crowds \cite{reiter1998crowds} is one of the first proposals to achieve user
anonymity. In it, a message is randomly forwarded between group members before
it reaches its destination. Therefore, none of the group members nor the end
recipient learn the actual source of the message. The Onion Router (TOR)
\cite{syverson2004tor} is another method that provides anonymous communication
through a ``circuit.'' Circuits are multi-hop encrypted communication channels
established using at least three TOR nodes. Theoretically, TOR guarantees
anonymity with respect to an adversary controlling, at most, two TOR nodes.
However, flawed TOR implementations can reduce its provided anonymity level
\cite{biryukov2013trawling}.

Hosts anonymity is not provided by ICING-based \neb. By inspecting packet
headers, eavesdroppers can easily determine a packet's source, as well as the
path it traversed. However, host anonymity can be achieved by replacing ICING
with TorIP \cite{Liu:2011:TII:2070562.2070576}, thus resulting in a level of
anonymity similar to that provided by TOR in today's Internet.

Unlike IP, NDN has some features that facilitate anonymous communication. A PIT
allows interest messages and content objects to only carry the requested content
name without any consumer-related information. However, Compagno et al. show in
\cite{geondn} that adversaries with enough knowledge of the network can determine
consumer's location. DiBenedetto et al. proposed {\sf AND\=aNA}
\cite{dibenedetto2011andana}, a tool that provides a level of anonymity similar
to TOR, while requiring only two intermediate nodes, instead of three.

MobilityFirst and XIA suffer from the same privacy and anonymity problems as IP.
Packets contain both source and destination GUIDs (or principal identifiers),
thus revealing the hosts involved. To make the matter worse, XIA packets path
can be revealed by inspecting their destination DAG addresses. This is because
such addresses might include (part of) the path to the destination, as described
in Section \ref{sec:xia}. Due to the communication model similarity of
these two architectures to IP, approaches developed to preserve users anonymity
in IP networks can be adopted. For instance, TOR can be used to protect
MobilityFirst and XIA host principals' anonymity. Preserving content principals'
anonymity can be achieved using a protocol similar to {\sf AND\=aNA} in NDN
\cite{dibenedetto2011andana}.

\subsection{Availability}\label{sec:network_dos_and_ddos}
Denial-of-Service (DoS) and Distributed DoS (DDoS) present a well known thread
against availability. In the following, we
discuss DoS and DDoS attacks that apply to the network layer of the current as
well as (some of) the future Internet architectures discussed above. We also
shed the light on new (and possibly more serious) type of attacks that the new
architectures allow. We exclude discussing the lack of availability due to
network misconfigurations, disasters, hardware and software faults, or any other
causes that are not a direct consequence of an attack. 

\newpar{Bandwidth depletion attacks}
The current Internet architecture is susceptible to bandwidth depletion attacks
\cite{specht2004distributed}. Their goal is to exhaust bandwidth of a specific
link. These attacks can be mounted in two ways: (1) distributed --
with packets sent at low rate by each attacking node, or (2) centralized -- a
single powerful adversary flooding the target link at high rate. Due to
today's high bandwidth and redundancy, centralized bandwidth depletion attacks
are harder to mount.

Several mitigation and prevention techniques have been proposed and implemented
in the current Internet. Some examples are: (1) tracing back traffic to the
source of the attack \cite{bellovin2003icmp,snoeren2001hash,savage2001network},
(2) distinguishing between legitimate and malicious traffic
\cite{anderson2004preventing,1301320}, (3) using puzzles to increase
the cost for adversaries trying to consume bandwidth
\cite{dean2001using,juels1999client}, and (4) using rate-limiting mechanisms for
traffic that causes congestion
\cite{ioannidis2002implementing,stoica2003core,mahajan2002controlling}.
However, none can effectively defeat this attack.

Bandwidth depletion in the data plane is harder to mount in \neb, because senders
(adversaries) must obtain consent of all nodes on a path before sending
packets. Thus, unauthorized packets will be dropped by adversary-facing routers.
Unfortunately, this only shifts the attack from the network layer to the consent
servers and causes negative effects on the network. The reason is because a
single consent server might be responsible for a large number of routers in its
domain. Thus, lowering its ability of issuing PoCs can disable all routers in
that domain.

NDN design is more resilient to bandwidth depletion attacks as compare to its
IP counterpart. Recall that NDN communication adheres to the {\em pull} model,
i.e., a content is forwarded only in response to a corresponding previous
interest. This model prevents adversaries from flooding the network with
unsolicited content. In addition, flooding the network with a large number of
interests to cause bandwidth depletion is not very effective due to the small
size of interest packets.

Since MobilityFirst and XIA use a communication model similar to IP, they both
are susceptible to bandwidth depletion attacks. Similar countermeasures
applied in IP networks can be adopted. However, they can only reduce the effect
of these attacks.

Nugraha et al. \cite{nugraha2014mutual} suggested integrating STRIDE with XIA to
protect against DoS attacks. STRIDE \cite{hsiao2013stride} is an architecture
resilient to bandwidth depletion (D)DoS attacks. It modifies SCION path
establishment to perform a tree-based bandwidth allocation. Whenever a trusted
domain root initiates the path establishment process, bandwidth is allocated as
the path is branching out as a tree from that root. This guarantees the required
bandwidth for benign flows. STRIDE also supports long-term static paths to
provide high available connectivity.

\newpar{Routers resource exhaustion}
Exhausting storage resources of routers is another target for adversaries.
NAT-enabled router might assemble
fragments in some scenarios. Reassembly buffers in these routers can be exploited
as follows. Each fragment includes a 16-bit field to indicate the size of the
original packet. Adversaries can send a single fragment with a large original
packet size and never send the rest of the fragments. This forces assembling
routers to allocate a buffer and wait for the rest of the packet to arrive. To
make the matter worse, adversaries can set the original packet size to its
maximum value, $64$KB, thus allocating maximum buffers.

Computation resources of routers is another victim of exhaustion. Since ICING's
design requires the extensive use of cryptographic operations, adversaries can
send a large number of packets to routers forcing them to perform all
verification operations described in Section \ref{sec:nebula}. Such attacks
have very low cost on adversaries since the latter can flood victim routers with 
packets carrying invalid (e.g., randomly generated) PoC and PoP values.

As mentioned above, PIT is one of the main router components that enables
content delivering without requiring any form of consumer source addresses. PIT
is also used for interest collapsing in order to reduce bandwidth due to a
burst of closely-spaced interests for the same content. However, the fact that
PIT is a limited and valuable resource makes it susceptible to malicious
exhaustion. Adversaries can send a large number of interests attempting to fill
the PIT. To avoid collapsing, such interests can refer to {\em nonsensical}
content. Once the PIT is full, a router can either: (1) drop incoming interests,
or (2) remove old PIT entries to make space for new ones. Both options, however,
can adversely impact past or future interests. This type of attack is called
Interest Flooding (IF)~\cite{gasti2013and,wahlisch2013backscatter}.

Unfortunately, there is no comprehensive remedy for IF attacks. Although several
countermeasures have been proposed, they are ineffective against smart
adversaries and only manage to lower the volume of IF attacks
\cite{CompagnoCGT13,6663516,al2015revisiting}.
One possible remedy for IF attacks is to eliminate the PIT -- its root
cause. For this reason, \cite{ghali2015living} suggests a modified
Content-Centric Networking (CCN) architecture without router PITs.

\newpar{Cache-related attacks}\label{sec:cache_attacks}
Regardless of the aforementioned benefits of in-network caching, it opens the
door for new types of DoS attacks that do not exist in today's Internet network
layer: content poisoning and cache pollution. In the following we describe the
resiliency of NDN, MobilityFirst, and XIA against these attacks. We exclude
\neb\ since its design does not provide in-network caching.

\paragraph{Content poisoning}
Content poisoning attacks occurs when adversaries
inject fake content into router caches. A fake content is not
generated by a benign producer and, consequently, does not satisfy user requests.
If such content is cached in routers, it is used to reply to future
benign user requests.

The fact that NDN adheres to the pull model makes it harder, but still feasible,
for adversaries to inject fake content into router caches. There are two methods
to achieve this:
\begin{itemize}
\item Reactive: the adversary \Adv\ is a node eavesdropping or controlling
  a link, e.g., an upstream malicious router. \Adv\ responds to
  interests on that links with a fake content that is cached in all downstream
  routers.
\item Proactive: this method involves \Adv\ that, anticipating demand for
  certain content, issues one or more bogus interests (perhaps from strategically
  placed zombie consumers), before genuine interests are issued. \Adv\ then
  replies with fake content (from a set of compromised routers or compromised
  producers) thus pre-poisoning the caches of all routers forwarding the bogus
  interests.
\end{itemize}
What facilitates this attack is the fact that content signature verification is
not mandatory by routers. This is because not only signature
verification is an expensive operation at line speed, variant trust models
adopted by different applications renders trusting public (verifying) keys
a challenge. Even if we assume that routers can know about all possible
trust models, some might incur heavy and expensive network operations. For
instance, PKI hierarchy requires routers to iteratively fetch and verify public
keys until a trusted anchor is reached. Such an impractical approach
significantly reduces the network performance.

Ghali et al. identifies in \cite{ghali2014network}, the root causes and proposes
a solution for content poisoning. The main idea is to have consumers and
producers collaborate in providing routers with enough trust contextual
information to perform a single signature verification.\footnote{We assume that
in the future, public key operations will be available at the hardware level.}
This process is formalized by a rule called the Interest Key Binding (IKB) rule.

An alternative solution is the use of SCNs. By definition, a SCN contains a value
that (uniquely) identifies the principal it is referring to. In case of content
principal in MobilityFirst and XIA, and content objects in NDN, SCNs contain
the hash of the data itself. When users request
content using SCNs, the network guarantees that the requested content will be
correctly delivered. As a result, MobilityFirst and XIA
eliminate the effects of content poisoning attacks by design. It is worth
mentioning that using SCNs does not prevent adversaries from injecting fake
content in router caches. Instead, it guarantees that benign users will not
receive such fake content.

\paragraph{Cache pollution}
Pollution is another type of (D)DoS attacks against router caches. In such
attacks, adversaries attempt to manipulate reference locality of caches, causing
incorrect decisions made by cache eviction strategies. This causes routers to
possibly evict popular content reducing the overall content distribution
performance. NDN, MobilityFirst, and XIA are all susceptible to this attack.

Conti et al. discuss this attack in \cite{conti2013lightweight}. It is shown that
with even limited adversarial resources, a highly effective cache pollution
attack can be mounted. In fact, even small cache locality manipulation can cause
a significant content distribution disruption \cite{deng2008pollution}. It is
also shown in \cite{conti2013lightweight} that launching pollution attacks on
large networks is relatively easy, and smart adversaries reduce the effectiveness
of proposed countermeasures.

Cache pollution attacks do not prevent users from retrieving the
requested data. Instead, they negatively effect the performance of content
distribution, and eliminate the benefits of in-network caching.

\section{Resolution Services Security}\label{sec:resolution_service}
Resolution service is a fundamental part of the current Internet architecture.
It maps human-readable names to routable network addresses. As mentioned above,
new Internet architectures also require similar resolution services to operate.
In this section we compare the security and privacy features of the various
resolution services proposed in FIA projects. We exclude \neb and XIA since they
do not propose a new resolution service and only exploit the existing DNS/DNSSEC.

We consider the following security features: trust, data origin authentication,
data integrity, peer entity authentication, authorization and access control,
accounting, data confidentiality and availability. We consciously exclude traffic
flow confidentiality and anonymous communication since we believe they should be
provided at the network layer.

\subsection{Trust}\label{subsec:rs_trust}
DNSSEC introduces the notion of trust into DNS. It considers authoritative
servers as trusted entities responsible of maintaining and securely providing
the correct mapping between human-readable domain names and corresponding IP
addresses. Recall that each DNSSEC server signs the resource records (name-to-IP
mappings) of its respective domain. Trustworthiness of such servers is ensured
by a chain-of-trust model that resembles the domains hierarchical organization.
The top-level domain resides at the root of this chain.

Similar to DNSSEC, NDNS applies the same notion. Authoritative server are trusted
entities and their trustworthiness is ensured by a similar chain-of-trust model.

GNRS, on the other hand, adopts a different approach. Every network
is responsible of providing signed GUID-NA mappings.
Thus, verifying these signatures ensures their validity. This also
prevents (compromised) GNRS from manipulating GUID-NA mappings without being
detected.

\subsection{Data-origin authentication and data integrity}
\label{subsec:rs_authentication_integrity}
DNSSEC provides data-origin authentication and data integrity by requiring: (1)
authoritative servers to sign each of their resource records, and (2) resolvers 
to verify the validity of these signatures and their corresponding public keys.
This prevents adversaries from injecting bogus data into the DNS system.

Signing every response resource record is an expensive operation that
authoritative server should not perform at run-time. Adversaries
can abuse such costly operation to launch (D)DoS attacks
against authoritative servers. To this end, resource record signatures should
always be generated in advance. While this method can be easily applied in case
resolvers asks for existing domain names, it does not work for a non-existing
domain names. DNS uses the NXDOMAIN resource records
to inform a resolver that the queried name does not exist. However, providing
data-origin authentication and integrity for NXDOMAIN resource records can not be
done by generating the signature in advance, because of the number of
possible non-existing names is infinite. To solve this
problem, DNSSEC introduces a new record type called the NextSECure (NSEC)
resource record. Specifically, assuming a canonical ordering of the domain names,
a NSEC record contains two consecutive existing names in the canonical ordering, thus
describing the gaps between them. Such records are signed and
used as authenticated denial of existence for non-existing names. Since NSEC
records contain existing names, their signatures can be calculated
{\em a priori}.

In NDNS, query responses are carried in content object payloads, thus data-origin
authentication and integrity is inherited from NDN. One
difference between DNSSEC and NDNS resides in the granularity of these
authentication and integrity guarantees. While DNSSEC can offer such security
properties per individual resource record or resource record set, the fact that
a NDNS record is carried in a content object can only guarantee the
authentication and integrity of said record. Although this is a clear restriction
in the flexibility and scalability of the protocol, it does not jeopardize its
security \cite{afanasyev2013addressing}.

In order to overcome DoS attacks due to requests for non-existing names, NDNS
adopts methods similar to DNSSEC. In particular, NDNS servers can sign the gaps
between existing names. Furthermore, Compagno et al. proposed in \cite{7288477}
the use of a Bloom filter \cite{bloom1970space} to defeat the
aforementioned DoS attacks. This, however, requires some changes in the NDN
architecture as well as the introduction of a new content type.

GNRS also provides data-origin authentication and integrity by means of GUID-NA
mapping signatures. The main different between GNRS and DNSSEC
is that the former does not assume that GNRS servers are trusted
\cite{liu2013secure}, while the latter requires all authoritative servers to be
trusted.

\subsection{Peer entity authentication}
In DNS and DNSSEC there is no entity authentication between a resolver and a DNS
server. Resolvers usually know the IP address of a DNS server which is used to
initiate queries. Usually, such IP address is manually configured
on the resolver or obtained through DHCP and considered valid. Using a TLS
connection between resolvers and DNS servers can provide entity authentication
\cite{bortzmeyer2015dns,zhu2014t}.

Authentication between DNS servers is obtained through Transaction signatures
(TSIG) \cite{vixie2000secret}. TSIG involves pairwise keys shared among DNS
servers and used to secure dynamic updates, zone transfers and recursive queries.
Moreover, in case of dynamic updates generated from DNS clients, a signature is
used to authenticate that  client and validate the update.

NDNS follows the same approach of DNS by not providing entity authentication.
However, the fact that both NDNS users and servers are regular NDN consumers and
producers, respectively, allows approaches similar to what is proposed in Section
\ref{subsec:peer-entity-authentication-network} to be adopted. Furthermore,
securing dynamic updates requires NDNS clients to have previously shared their
certificate with the NDNS servers. Signing the updates will then authenticate
them.

Unlike DNS and NDNS, GNRS authenticates every client issuing requests (query,
update and delete) by retrieving the corresponding certificate, from NCS, and
verifying the GUID authenticity. GNRS clients can ensure server authentication
by performing similar steps, requesting certificates from NCS and verifying
servers identities.

\subsection{Authorization and access control}\label{subsec:rs_access_control}
Both DNS/DNSSEC and NDNS do not provide any form of access control. All resource
records are publicly available to every host in the network. However,
authorization and access control is provided for dynamic updates.

GNRS does not follow the same trend and consider access control a
crucial part of its design. Specifically, GNRS stores a set of access control
policies along with GUID-NA mappings. Such policies regulate access to particular
GNRS resources by specifying read and write permissions which blacklist and
whitelist certain user GUIDs.

\subsection{Accountability}\label{subsec:rs_accounting}
DNS/DNSSEC guarantees accountability only for secure DNS dynamic updates
\cite{wellington2000secure}. This is because such requests must in fact be signed
by their originators. Also, NDNS does not provide any mechanism for
accountability. The fact that NDNS users and servers are consumers and
producers allows the adoption of similar approaches described in Section
\ref{sec:accountability}.

GNRS uses a different approach and mandates GNRS clients to sign every request.
By doing so, accountability is provided for all insert, update and query
requests. 

\subsection{Data confidentiality}\label{subsec:rs_confidentiality}
Neither DNS nor DNSSEC provide confidentiality. Queries and resource records are
never encrypted by authoritative servers and are always exchanged in cleartext.
One way of achieving confidentiality in DNS/DNSSEC is to establish a TLS
session between resolvers and authoritative servers
\cite{bortzmeyer2015dns,zhu2014t}.

Similarly, both NDNS and GNRS designs do not take confidentiality into
consideration. Fortunately, similar approaches to using TLS channels can
be adopted. This feature is important to provide private communication

\subsection{Availability}\label{subsec:rs_ddos}
As a public available service, DNS is subject to
(D)DoS attacks which jeopardize its availability. In particular, adversaries
can flood authoritative servers with a large 
number of query requests to exhaust their resources.\footnote{The use of secure
dynamic updates involving asymmetric encryption increases the effect of the
attack.} Although the use of DNS caching and redundancy servers reduce the effect
of DDoS, a number of such attacks have been successfully directed against root
and top-level DNS servers in past years \cite{dnsddos1,dnsddos2,dnsddos3}.
Many solutions have been presented in past years that either: (1)
require some changes in the DNS protocol \cite{cox2002serving,deegan2005main,
ramasubramanian2004design,pappas2007enhancing}, or (2) propose new resolution
services \cite{yang2004hours,handley2005case}. Nowadays, DNS uses a single
approach that does not require any modification to its architecture, which is the
adoption of ``Anycast'' \cite{abley2006operation}. In this case, a single DNS
server is replicated in different geographically locations among several ASes.
Therefore, routing protocols forward DNS requests to the nearest server that can
satisfy them \cite{conradtowards}. However, this approach can not achieve an
efficient load balancing because it does not consider replica workloads and
network traffic.

DNS resolvers (not implementing DNSSEC protocol) can be the target of cache
poisoning attacks. This attack is similar, in concept, to content poisoning
described in Section \ref{sec:cache_attacks}. In DNS cache poisoning attacks,
the goal of the adversary is to inject spoofed responses (name-IP mappings) in
the resolver caches \cite{rfc3833}. Injecting false DNS responses can be achieved
using a man-in-the-middle attack in which the adversary satisfies requests
with false DNS responses \cite{rfc3833}. The introduction of data-origin
authentication in DNSSEC allows resolvers to verify the origin of data in
DNS response, thus eliminating this attack.

NDNS follows the same hierarchical design of DNS. In principles NDNS
authoritative servers seems to offer the same level of resilience to DDoS as DNS
authoritative servers. Furthermore, NDNS envisions the use of a set of secondary
servers to balance the workload, which was proven to be not effective. The same
anycast approach used in DNS could be employed in NDNS. Such forwarding
strategy must be implemented by NDN routers. Moreover, NDNS is susceptible to
content poisoning attacks. Fortunately, the same countermeasures described in
\ref{sec:cache_attacks} can be effective.

GNRS design appears to be more resilient to DDoS than DNS design. In fact, GNRS
does not adopt a hierarchical structure, instead it distributes the GUID-NA
mappings among a number of replicas using Distributed Hashtables (DHT) maintained
by all the ASes in the Internet. This allows GNRS to easily scale and distribute a
DDoS attack.

\section{Lessons learned and Future Directions}\label{sec:summary}
In this section we summarize all the security and privacy features provided by
the network layer of the FIA architectures and their corresponding resolution
services. We also highlight the missing features and delineate a direction for
future work on security and privacy in both network layer and resolution
services.

\subsection{Network Layer}
The strong requirement to include security and privacy by design seems to have
been only partially reached by FIA funded
projects. Table~\ref{tb:network_comparison} summarizes the security and privacy
features provided by each architecture.

\begin{table*}[t]
  \def\arraystretch{1.3}
  \centering
  \begin{tabular}{|l|p{2cm}|p{2cm}|p{2cm}|p{2cm}|}
    \hline
    \multirow{2}{*}{\bf Security and Privacy Features} & \multicolumn{4}{c|}{{\bf Network layers}}\\
    \cline{2-5}
    & \multicolumn{1}{c|}{\bf\neb} & \multicolumn{1}{c|}{\bf NDN} 
    & \multicolumn{1}{c|}{\bf MobilityFirst} & \multicolumn{1}{c|}{\bf XIA} \\
    \hline
    Trust 
    & \multicolumn{1}{c|}{\cmark} & \multicolumn{1}{c|}{\cmark} 
    & \multicolumn{1}{c|}{\cmark} & \multicolumn{1}{c|}{\cmark} \\
    \hline
    Data Origin Authentication 
    & \multicolumn{1}{c|}{$\circledcirc$} & \multicolumn{1}{c|}{\cmark} 
    & \multicolumn{1}{c|}{\xmark} & \multicolumn{1}{c|}{\xmark} \\
    \hline
    Peer entity Authentication 
    & \multicolumn{1}{c|}{$\circledcirc$} & \multicolumn{1}{c|}{$\circledcirc$} 
    & \multicolumn{1}{c|}{$\circledcirc$} & \multicolumn{1}{c|}{$\circledcirc$} \\
    \hline
    Data Integrity 
    & \multicolumn{1}{c|}{$\circledcirc$} & \multicolumn{1}{c|}{\cmark} 
    & \multicolumn{1}{c|}{\xmark} & \multicolumn{1}{c|}{\xmark}\\
    \hline
    Authorization and Access Control
    & \multicolumn{1}{c|}{\cmark} & \multicolumn{1}{c|}{$\circledcirc$} 
    & \multicolumn{1}{c|}{$\circledcirc$} & \multicolumn{1}{c|}{$\circledcirc$} \\
    \hline
    Accountability 
    & \multicolumn{1}{c|}{\cmark} & \multicolumn{1}{c|}{$\circledcirc$} 
    & \multicolumn{1}{c|}{$\circledcirc$} & \multicolumn{1}{c|}{$\circledcirc$} \\
    \hline
    Data Confidentiality  
    & \multicolumn{1}{c|}{\xmark} & \multicolumn{1}{c|}{\cmark} 
    & \multicolumn{1}{c|}{\xmark} & \multicolumn{1}{c|}{\xmark} \\
    \hline
    Traffic Flow Confidentiality 
    & \multicolumn{1}{c|}{\xmark} & \multicolumn{1}{c|}{\xmark} 
    & \multicolumn{1}{c|}{\xmark} & \multicolumn{1}{c|}{\xmark} \\
    \hline
    Anonymous Communication 
    & \multicolumn{1}{c|}{\xmark} & \multicolumn{1}{c|}{\xmark} 
    & \multicolumn{1}{c|}{\xmark} & \multicolumn{1}{c|}{\xmark} \\
    \hline
    Availability 
    & \multicolumn{1}{c|}{$\circledcirc$} & \multicolumn{1}{c|}{$\circledcirc$} 
    & \multicolumn{1}{c|}{$\circledcirc$} & \multicolumn{1}{c|}{$\circledcirc$} \\
    \hline
  \end{tabular}
  \vspace{0.2cm}
  \caption{Network Layer Security and Privacy Comparison. \cmark indicates that
    the feature is fully considered and available in the architecture. \xmark
    indicates that the feature is not available and not considered in the
    architecture. $\circledcirc$ indicates that the feature is partially
    available or the architecture provides some mechanisms to facilitate
    implementing the feature by the application.}
  \label{tb:network_comparison}
\end{table*}

\newpar{\neb} 
At its current state, \neb\ includes trust, data origin authentication, peer
entity authentication, data integrity, authorization and access control,
accountability, and availability features. All of them are provided between
senders and routers implementing ICING, expect for peer entity authentication
that is guaranteed between senders and consent servers during the retrieval of
the PoCs. With respect to IP and IPsec, Nebula certainly increases the security
of intra-domain communication. However, it lacks to consider inter-domain and
end-to-end communication security. While current approaches can be adopted,
e.g., establishing an IPsec-like secure communication between two end hosts,
researcher should investigate these two directions further.

Availability in Nebula is provided by path verification mechanism which prevents
any adversary from sending unrequested data to perform bandwidth depletion
attacks. However, ICING nodes and consent servers could be the target of a DoS
attack, due to the heavy use of cryptographic operations.  Consent servers can
be another target for DoS attacks. In particular, an adversary can flood a
server with an abnormal amount of request. If the target server controls a large
number of routers, the attack can effectively disable all of them.  We believe
that such attack deserves further investigation.

Other analyzed features, i.e., data confidentiality, traffic flow
confidentiality and anonymous communication, are not currently provided. The
first two can be easily implemented by applying the same well-known technique
used in IP/IPsec. However, \neb\ should investigate and make these features
available by design before any adoption in the real world. Furthermore,
anonymous communications contradict with the current design of ICING which
requires path establishment for each communication. Therefore, further
investigation is required by the research community to provide anonymity in
\neb.

\newpar{NDN} 
This architecture replaces hosts as primary entity in the network with
content. This affect the security features provided at network layer by focusing
them on content rather than end-hosts. Currently, NDN provides: trust, data
origin authentication, data integrity and data confidentiality. Security
features like peer entity authentication and accountability are available only
when derived from data origin authentication for both interest and content.
Ubiquitous caching further complicates achieving these two features. In fact,
when a consumer obtains a content from an intermediate router's cache, there is
currently no way to provide peer entity authentication between the consumer and
the router providing the content. Similarly, accountability for content refers
to the producer generating that content and not to any intermediate router
serving it from its cache.

Availability is a security feature in which NDN provides some improvements as
compared to IP/IPsec but at the same time it opens the door for new types of
attacks. While the pull model communication prevents any adversary from flooding
a host with non-requested content, adversaries can exhaust router states (i.e.,
PIT and CS), decreasing the performance of a network. Even through several
countermeasures have been proposed, none of them has been chosen and implemented
as part of the architecture.

The remaining security features: authorization and access control, traffic flow
confidentiality and anonymous communication are not provided by the network
layer. Even if ACL can be implemented, NDN chooses to delegate access control to
the application layer by encrypting content and distributing the decryption keys
only to authorized consumers. Traffic flow confidentiality and anonymous
communication are not available but existing approaches designed for IP and
IPsec can be applied without any modification of the NDN architecture.

\newpar{MobilityFirst} 
The security features provided by this architecture combine in principle the
end-to-end hosts approach and the content-based approach of IP/IPsec and NDN,
respectively. However, at the current state, MobilityFirst seems to fully
provide only few security features, i.e., trust, authorization and access
control.  The adoption of self-certifying names to address hosts facilitates the
implementation of data origin authentication, peer entity authentication, and
accountability. Peer entity authentication can be achieved involving a simple
challenge-response protocol between the two peers. While this guarantees
end-to-end accountability, it is not enough to provide network
accountability. In order to achieve that, edge routers should prevent address
spoofing.

Data integrity and data confidentiality are currently not provided at network
layer and it is not clear if MobilityFirst approach is to let applications deal
with them. Traffic flow confidentiality and anonymous communication are also not
provided. Existing approaches designed for IP/IPsec can be easily
imported. However, we encourage a deeper investigation in order to design new
techniques able to exploit the peculiarity of the new architecture.

The current design of MobilityFirst can help against content poisoning attacks
(i.e., the use of SCN prevents a benign user from receiving fake
content). Meanwhile, MobilityFirst does not seem to have fully considered other
network attacks such as bandwidth depletion attacks and cache pollution
attacks. Existing solutions can be applied to mitigate these problems. We
believe a further investigation is worth being conducted in order to provide
some new and architecture-related countermeasures.

\newpar{XIA} 
The main goal of this architecture is to support communication between multiple
and different principals. XIA security approach extends, de-facto, the
MobilyFirst approach in which security is provided and designed for two
principals: content and hosts. XIA does not limit the number of principals but
instead provides the freedom to design new ones. To this end, XIA security is
based on each principal intrinsic security feature.

The current state of XIA offers the same security and privacy features as
MobilityFirst. This is due to their similar approach in addressing principals
using SCNs.

\subsection{Resolution Services}
Resolution services are of a primary importance in both the current Internet
architecture and the new FIA proposals. In Nebula, the resolution service is
used by the NVENT to perform path discovery. In NDN, the same service provides
the mapping between namespaces and the corresponding security information. In
MobilityFirst, the resolution service is actively involved in any communication
guaranteeing the binding between GUIDs and NAs. Finally, XIA's DNS-based
resolution service is used for the address/path resolution.

Although all architectures requires a resolution service, only NDN and
MobilityFirst are actually investigating their own proposal. In
Table~\ref{tb:resolution_service_comparison} we summarize the security and
privacy features provided by NDNS and GNRS.

\begin{table*}[t]
  \def\arraystretch{1.2}
  \centering
  \begin{tabular}{|l|p{2cm}|p{2cm}|}
    \hline
    \multirow{2}{*}{\bf Security and Privacy Features} 
    & \multicolumn{2}{c|}{{\bf Resolution Services}}\\
    \cline{2-3}
    & \multicolumn{1}{c|}{\bf NDNS} & \multicolumn{1}{c|}{\bf GNRS} 
   \\
    \hline
    Trust  
    & \multicolumn{1}{c|}{\cmark} & \multicolumn{1}{c|}{\cmark} \\
    \hline
    Data Origin Authentication 
    & \multicolumn{1}{c|}{\cmark} & \multicolumn{1}{c|}{\cmark} \\
    \hline
    Peer entity Authentication 
    & \multicolumn{1}{c|}{\xmark} & \multicolumn{1}{c|}{\cmark} \\
    \hline
    Data Integrity 
    & \multicolumn{1}{c|}{\cmark} & \multicolumn{1}{c|}{\cmark}\\
    \hline
    Authorization and Access Control
    & \multicolumn{1}{c|}{\xmark} & \multicolumn{1}{c|}{\cmark} \\
    \hline
    Accountability 
    & \multicolumn{1}{c|}{$\circledcirc$} & \multicolumn{1}{c|}{\cmark} \\
    \hline
    Data Confidentiality  
    & \multicolumn{1}{c|}{\xmark} & \multicolumn{1}{c|}{\xmark} \\
    \hline
    Availability 
    & \multicolumn{1}{c|}{$\circledcirc$} & \multicolumn{1}{c|}{\cmark} \\
    \hline
  \end{tabular}
  \vspace{0.2cm}
  \caption{Resolution Services Security and Privacy Comparison. \cmark indicates
    that the feature is fully considered and available in the architecture.
    \xmark indicates that the feature is not available and not considered in the
    architecture. $\circledcirc$ indicates that the feature is partially
    available or the architecture provides some mechanisms to facilitate
    implementing the feature by the application.}
  \label{tb:resolution_service_comparison}
\end{table*}

\newpar{NDNS}
NDNS design reflects in many part the design of DNS without bringing too much
novelty. NDNS involves the same notion of trust of DNSSEC in which servers are
trusted entities. Moreover, NDNS queries and responses provide: data origin
authentication, data integrity, accountability only in case of dynamic updates,
and availability. The last feature is provided by server replication.  However,
while in DNSSEC the ``Anycast'' technique is used to choose the closes server to
the resolver, in NDNS the network must be aware of all servers and implement
specific forwarding strategies to balance the requests among them, which adds
more complexities.

Peer entity authentication, authorization and access control, and data
confidentiality are not provided. We believe that NDNS deserves further
investigation to provide better availability, as well as the missing security
features.

\newpar{GNRS}
The GNRS design is completely different from the current DNS and DNSSEC. In fact
MobilityFirst assumes a flat name structure which forces GNRS to assume a flat
structure for its servers. This different organization has some side effects on
the provided security features: servers are not assumed to be trusted entities
and, data origin authentication is provided by the owner of the GUID and not by
servers. GNRS also introduces authorization and access control of its stored
information and provides accountability and peer entity authentication in every
query and response.  Finally, GNRS is more robust to DoS attacks than
DNS. First, compromising one server does not affect others.  Second, GNRS is
designed to easily adapt to network changes and to balance GUID-NA mapping among
many replicas based on requests locality.

We believe that GNRS introduces good security improvement with respect to the
current DNS and DNSSEC. However, the only missing feature is data
confidentiality.

\section{Conclusion}\label{sec:conclusions}
Despite the unquestionable success of the current IP-based Internet
architecture, the lack of security considerations in its design lead to many
severe security breaches and privacy leakages for many years. One goal of future
Internet architectures is to include security and privacy features that are
missing in the design of the current Internet.
  
In this survey, we provided a thorough analysis of security and privacy features
currently supported by the network layer of four architectures involved in the
FIA program: Nebula, NDN, MobilityFirst and XIA. We focused on features we
believe should be available at the network layer in order to achieve secure and
private communication. We considered IP/IPsec as a point of reference in our
analysis.  Moreover, we included resolution services provided by the different
architectures since each one can not abstain from relying on such a service.

\balance
\bibliographystyle{IEEEtran}
\bibliography{IEEEabrv,references}

\end{document}